\documentclass[iop]{emulateapj}

\slugcomment{Accepted for publication in Astrophysical Journal on December 8, 2015}

\defcitealias{2013PASJ...65..100I}{I13} 

\usepackage{color}
\usepackage{apjfonts}
\usepackage{natbib}
\usepackage[dvipdfmx,colorlinks=true,linkcolor=blue,filecolor=blue,urlcolor=blue,citecolor=blue]{hyperref}
\hypersetup
{
	pdfauthor={T. Izumi et al.},
	pdftitle={Izumi2014}
}

\shorttitle{Submm-HCN diagram for an energy diagnostics in galaxies}
\shortauthors{T. Izumi et al.}

\begin{document}

\title{Submillimeter-HCN diagram for an energy diagnostics in the centers of galaxies}


\author{Takuma Izumi\altaffilmark{1}, 
Kotaro Kohno\altaffilmark{1,2}, 
Susanne Aalto\altaffilmark{3}, 
Daniel Espada\altaffilmark{4,5,6},
Kambiz Fathi\altaffilmark{7}, 
Nanase Harada\altaffilmark{8}, 
Bunyo Hatsukade\altaffilmark{5}, 
Pei-Ying Hsieh\altaffilmark{8,9}, 
Masatoshi Imanishi\altaffilmark{10,5,6}, 
Melanie Krips\altaffilmark{11}, 
Sergio Mart{\'{i}}n\altaffilmark{11,12,4}, 
Satoki Matsushita\altaffilmark{8}, 
David S. Meier\altaffilmark{13}, 
Naomasa Nakai\altaffilmark{14}, 
Kouichiro Nakanishi\altaffilmark{4,5,6}, 
Eva Schinnerer\altaffilmark{15}, 
Kartik Sheth\altaffilmark{16}, 
Yuichi Terashima\altaffilmark{17}, 
and Jean L. Turner\altaffilmark{18} 
}

\affil{${}^{1}$\ Institute of Astronomy, School of Science, The University of Tokyo, 2-21-1 Osawa, Mitaka, Tokyo 181-0015, Japan; \href{mailto:takumaizumi@ioa.s.u-tokyo.ac.jp}{takumaizumi@ioa.s.u-tokyo.ac.jp}}
\affil{${}^{2}$\ Research Center for the Early Universe, The University of Tokyo, 7-3-1 Hongo, Bunkyo, Tokyo 113-0033} 
\affil{${}^{3}$\ Department of Earth and Space Sciences, Chalmers University of Technology, Onsala Observatory, 439 94 Onsala, Sweden}
\affil{${}^{4}$\ Joint ALMA Observatory, Alonso de C{\'{o}}rdova, 3107, Vitacura, Santiago 763-0355, Chile} 
\affil{${}^{5}$\ National Astronomical Observatory of Japan, 2-21-1 Osawa, Mitaka, Tokyo 181-8588, Japan}
\affil{${}^{6}$\ SOKENDAI (The Graduate University for Advanced Studies), 2-21-1 Osawa, Mitaka, Tokyo 181-8588, Japan}
\affil{${}^{7}$\ Stockholm Observatory, Department of Astronomy, Stockholm University, AlbaNova Centre, 106 91 Stockholm, Sweden}
\affil{${}^{8}$\ Academia Sinica, Institute of Astronomy \& Astrophysics, P.O. Box 23-141, Taipei 10617, Taiwan} 
\affil{${}^{9}$\ Institute of Astronomy, National Central University, No. 300, Jhongda Road, Jhongli City, Taoyuan County 32001, Taiwan, Republic of China} 
\affil{${}^{10}$\ Subaru Telescope, NAOJ, 650 North A'ohoku Place, Hilo, HI 96720, USA} 
\affil{${}^{11}$\ Institut de Radio Astronomie Millim{\'{e}}trique, 300 rue de la Piscine, Domaine Universitaire, 38406 St. Martin d'H{\`{e}}res, France}
\affil{${}^{12}$\ European Southern Observatory, Alonso de C{\'{o}}rdova, 3107, Vitacura, Santiago, Chile} 
\affil{${}^{13}$\ Department of Physics, New Mexico Institute of Mining and Technology, 801 Leroy Place, Soccoro, NM 87801, USA}
\affil{${}^{14}$\ Department of Physics, Faculty of Pure and Applied Sciences, University of Tsukuba, 1-1-1 Ten-nodai, Tsukuba, Ibaraki 305-8571, Japan}
\affil{${}^{15}$\ Max Planck Institute for Astronomy, K{\"{o}}nigstuhl 17, Heidelberg 69117, Germany}
\affil{${}^{16}$\ NASA, 300 E Street SW, Washington, DC 20546, USA}
\affil{${}^{17}$\ Department of Physics, Ehime University, 2-5 Bunkyo-cho, Matsuyama, Ehime 790-8577, Japan}
\affil{${}^{18}$\ Department of Physics and Astronomy, UCLA, 430 Portola Plaza, Los Angeles, CA 90095-1547, USA}


\begin{abstract}
Compiling data from literature and the ALMA archive, we show enhanced 
HCN(4-3)/HCO$^+$(4-3) and/or HCN(4-3)/CS(7-6) integrated intensity ratios in 
circumnuclear molecular gas around active galactic nuclei (AGNs) 
compared to those in starburst (SB) galaxies (submillimeter HCN-enhancement). 
The number of sample galaxies is significantly increased from our previous work. 
We expect this feature could potentially  
be an extinction-free energy diagnostic tool of nuclear regions of galaxies. 
Non-LTE radiative transfer modelings of the above molecular emission lines 
involving both collisional and radiative excitation, as well as a photon trapping effect 
were conducted to investigate the cause of the high line ratios in AGNs. 
As a result, we found that enhanced abundance ratios of HCN-to-HCO$^+$ and HCN-to-CS 
in AGNs as compared to SB galaxies by a factor of a few to even $\ga$ 10 
is a plausible explanation for the submillimeter HCN-enhancement. 
However, a counter argument of a systematically higher 
gas density in AGNs than in SB galaxies can also be a plausible scenario. 
Although we could not fully discriminate these two scenarios at this moment 
due to insufficient amount of multi-transition, multi-species data, 
the former scenario equivalently claims for abnormal chemical composition in AGNs. 
Regarding the actual mechanism to realize the composition, 
we suggest it is difficult with conventional gas phase X-ray dominated region (XDR) 
ionization models to reproduce the observed high line ratios. 
We might have to take into account other mechanisms such as 
neutral-neutral reactions that are efficiently activated at high temperature environments 
and/or mechanically heated regions to further understand the high line ratios in AGNs. 
\end{abstract}


\section{Introduction}\label{sec1} 
The dense molecular medium in the centers of galaxies plays a key role in their evolution 
because it is the reservoir of fuel for active galactic nuclei (AGNs), 
as well as the site of massive star formation (starburst = SB). 
On the other hand, these heating sources can conversely 
alter the chemical composition of their surrounding medium in radiative and/or mechanical ways. 
Thus, investigating such dense molecular gas can be essential 
to study the nature of the nuclear region accompanying these activities. 

Recent chemical modelings of interstellar medium (ISM) predict 
that the various heating mechanisms will produce different signatures in molecular gas composition. 
For example, intense UV radiation from massive stars forms photodissociation regions (PDRs) around them, 
and X-ray dominated regions (XDRs), which are larger in volume than PDRs 
due to the higher penetrating capability of the X-ray, are formed at the vicinity of AGNs 
(e.g., \citealt{1996ApJ...466..561M,1997ARA&A..35..179H,1999RvMP...71..173H,
2005A&A...436..397M,2006ApJ...650L.103M,2007A&A...461..793M,2008ApJ...676..978B,2009ApJ...696.1466B}). 
Cosmic rays from frequent supernovae (SNe) and the injection of mechanical energy 
due to SNe or AGN jet/outflow (mechanical heating) are also important in shaping chemical composition 
(e.g., \citealt{2011A&A...525A.119M,2011MNRAS.414.1583B,2008A&A...488L...5L,
2012A&A...542A..65K,2015A&A...574A.127K,2014A&A...564A.126R,2014A&A...568A..90R,2015ApJ...799...26M}). 
In addition to the gas phase reactions, inclusion of dust grain chemistry 
will influence the chemistry as well (e.g., \citealt{2005IAUS..231..237C,2008ApJ...682..283G}). 
It is noteworthy that the temperature of the gas is essentially important 
for the efficiency of chemical reactions, 
especially at the vicinity of powerful heating sources 
(e.g., \citealt{2001ApJ...546..324R,2004A&A...414..409N,2010ApJ...721.1570H,2013ApJ...765..108H}); 
for example, under the high temperature conditions, we can expect 
an enhanced abundance of HCN due to the activated neutral-neutral reactions from CN. 

Based on the above descriptions, we can deduce that it would be possible to construct 
a diagnostic method of the dominant energy source in galaxies such as an AGN and a SB 
by utilizing the potential chemical differences between them. 
To do so, millimeter/submillimeter spectroscopic observations are of great importance 
because these wavelengths do not suffer from dust extinction, 
which is critical to probe dusty nuclear regions. 
In addition, a high spatial and spectral resolution provided by millimeter/submillimeter interferometers 
can also give us critical information on gas kinematics in the central $\la$ 100 pc scale region, 
which is necessary to study feeding mechanisms of 
central supermassive black holes (e.g., \citealt{2013ApJ...770L..27F}). 

With these things in mind, many key molecules have been raised so far 
as useful observational diagnostic tools. 
Among them, an enhanced intensity of HCN(1-0), 
whose critical density ($n_{\rm cr}$) for collisional excitation is $n_{\rm H_2} \sim 10^{4-5}$ cm$^{-3}$, 
with respect to HCO$^+$(1-0) and/or CO(1-0) has been proposed as a unique feature to AGNs 
(e.g., \citealt{1993ApJ...418L..13J,1994ApJ...426L..77T,1994ApJ...436L.131S,2001ASPC..249..672K,
2004A&A...419..897U,2005AIPC..783..203K,2007AJ....134.2366I,2008ApJ...677..262K,2012A&A...537A.133D}). 
Using these line ratios, \citet{2001ASPC..249..672K} constructed 
a potential diagnostic diagram ({\it{mm-HCN diagram}}; see also \citealt{2005AIPC..783..203K}). 
But there are some counter arguments to this diagnostics that high HCN(1-0)/HCO$^+$(1-0) 
ratios are observed in non-AGNs (\citealt{2011A&A...528A..30C,2011AJ....141...38S}), 
as well as low HCN(1-0)/HCO$^+$(1-0) in AGNs (e.g., \citealt{2012MNRAS.424.1963S}). 
The latter inconsistency could be due to 
strong spectral contamination from the likely coexisting SB activities 
that dilutes emission from an AGN-influenced region such as an XDR (see also Section \ref{sec4.6}). 
Furthermore, the cause of the enhancement has not been clear 
because many different mechanisms can contribute to this enhancement; 
higher gas opacities, densities, and temperatures (excitation conditions), 
and/or abundance variations caused by different heating mechanisms. 
A non-collisional excitation, such as infrared (IR)-pumping 
caused by the re-radiation from UV/X-ray heated dust, 
could also be important (e.g., \citealt{2002A&A...381..783A,2007A&A...464..193A,
2006ApJ...640L.135G,2010ApJ...725L.228S,2013AJ....146...91I,2015ApJ...799...26M}), 
especially in (ultra) luminous infrared galaxies = (U)LIRGs. 

Similar to the above mentioned $J$ = 1--0 transitions, 
\citet{2013PASJ...65..100I} (hereafter \citetalias{2013PASJ...65..100I}) found that 
HCN(4-3)/HCO$^+$(4-3) and HCN(4-3)/CS(7-6) integrated intensity ratios 
seem to be higher in AGNs than in SB galaxies ({\it{submillimeter HCN-enhancement}}), 
and proposed a diagnostic diagram hereafter referred to as a $``${{\it{submm-HCN diagram}}}$"$ using these line ratios. 
One advantage of the submm-HCN diagram over the previous diagnostics using HCN(1-0) line would be 
that higher angular resolution is easily achievable at higher-$J$ compared to $J$ = 1--0 transitions, 
which is essentially important to exclude contamination from SB activity to the line emission from AGN-heated gas. 
Moreover, it is much more applicable to high-redshift galaxies by using high-$J$ lines 
because submillimeter lines can be covered by {\it{Atacama Large Millimeter/submillimeter Array (ALMA)}} 
up to, e.g., $z$ $\sim$ 3 for the case of $J$ = 4--3 transitions of HCN and HCO$^+$. 
These lines can be simultaneously observed with ALMA, which is necessary to obtain accurate line ratios, 
in terms of both little systematic flux uncertainty and, to a lesser extent, differences in the $uv$ coverage. 
In addition, since these transitions have orders of magnitude higher critical densities 
($n_{\rm H_2}$ $\sim$ 10$^{6-7}$ cm$^{-3}$) than the $J$ = 1--0 transitions ($n_{\rm H_2}$ $\sim$ 10$^{4-5}$ cm$^{-3}$), 
they are less contaminated by the foreground and/or disk emission, 
thus suitable to probe the densest gas in the obscured nuclear regions of galaxies. 

However, despite the above advantages, 
the proposed diagram of \citetalias{2013PASJ...65..100I} 
was very tentative as (1) it was based on as small as 5 galaxy sample, 
and (2) they mixed galaxies observed at a wide range of spatial resolutions ($\sim$ 100 pc to 1 kpc), 
which would combine flux contributions from various components. 
In order to assess the potential of the submm-HCN diagram, 
we need high spatial resolution observations allowing us 
to isolate the AGN emission from likely contamination due to co-existing SB, 
together with SB samples observed at matched spatial resolutions. 
Indeed, \citet{2010A&A...513A...7S} predicted that a distance out to which AGN-heating 
with X-ray luminosity of $\sim$ 10$^{43}$ erg s$^{-1}$ (a typical value for luminous Seyfert galaxies) dominates in dense gas 
($n_{\rm H_2}$ $\sim$ 10$^{4-5}$ cm$^{-3}$) with a soft UV 
radiation field of $\sim$ 100 $G_0$\footnote{1 $G_0$ = 1.6 $\times$ 10$^{-3}$ erg s$^{-1}$ cm$^{-2}$} 
is only $\la$ 100 pc (see also Appendix-A of \citealt{2015ApJ...811...39I}). 
Furthermore, high resolution observations of the nuclear regions of nearby Seyfert galaxies 
revealed that there sometimes is a hundreds-to-kpc scale ring-like SB region, 
which surrounds 100 pc scale dense molecular gas concentrations at the nucleus 
both in luminous AGNs (e.g., \citealt{2014A&A...567A.125G,2015ApJ...811...39I}) 
and low-luminosity AGNs (\citetalias{2013PASJ...65..100I}; \citealt{2015A&A...573A.116M}). 
We hereafter call the latter kind of (100 pc scale) 
central molecular concentration a {\it{circumnuclear disk (CND)}} in general. 

In this paper, we compile literature and archival data of 
HCN(4-3), HCO$^+$(4-3), and CS(7-6) emission lines of various AGNs and SB galaxies 
to improve the statistics of the submm-HCN diagram, 
and to explore the cause of the HCN-enhancement. 
Section \ref{sec2} describes the collected data. 
An up-dated submm-HCN diagram is shown in Section \ref{sec3}. 
Section \ref{sec4} presents simple non-local thermodynamical equilibrium (non-LTE) 
radiative transfer models involving HCN(4-3), HCO$^+$(4-3), and CS(7-6). 
We examine the impact of both excitation and molecular abundance on the line ratios there. 
In Section \ref{sec5}, we discuss possible chemical scenarios 
to realize the potential abundance variation suggested in Section \ref{sec4}. 
At last, our main conclusions of this work are summarized in Section \ref{sec6}.

\section{Data}\label{sec2}
In this work, we first compiled interferometric data of 
the target emission lines of extragalactic objects from literatures. 
We refer to data obtained with spatial resolutions better than 500 pc as the {\it{high resolution sample}}. 
The threshold resolution of 500 pc is large enough to fully encompass the typical size of CNDs in nearby galaxies, 
as well as small enough to exclude line emission from 
non-CND components such as circumnuclear SB rings in most cases. 
These data are further compared with rather lower resolution interferometric data (spatial resolution $>$ 500 pc) 
and single dish data (typical spatial resolution $>$ 1 kpc) 
with firm detections of emission lines ($>$ 5$\sigma$). 
These data are called {\it{low resolution sample}}, and will be used to investigate 
the impact of different spatial resolution on our diagnostics. 
As a result, we compiled line emission data of NGC 1068, NGC 1097, NGC 1365, 
NGC 4945, NGC 7469, NGC 4418, IRAS12127-1412, M82, NGC 253, NGC 1614, 
NGC 3256, NGC 3628, NGC 7552, IRAS 13242-5713, N113 (LMC), and N159 (LMC). 
Moreover, in the case with sufficiently high resolution data, 
we measured line ratios at different representative positions within the same galaxy. 
The name of each position such as NGC 1097 (AGN) and NGC 1097 (SB ring) is used hereafter. 
Note that we classify the data of NGC 4945 obtained with APEX $\sim$ 450 pc (18$\arcsec$) aperture into the low resolution sample, 
because it hosts a relatively compact circumnuclear SB ring with a radius of 2$''$.5 ($\sim$ 60 pc) inside the beam. 
On the other hand, the spatial resolution of the two LMC objects are orders of magnitude higher 
than the other high resolution sample due to their proximity. 
We nevertheless keep using their ratios considering the rarity of high resolution 
extragalactic measurements of the emission lines used in this work. 
We emphasize that excluding these LMC objects does not change our conclusion at all. 

Then, the total number of data points is 16 and 9 
for the high and the low resolution sample, respectively. 
Hence, we improved the statistics significantly (a factor of 5 for the combined sample) from \citetalias{2013PASJ...65..100I}. 
The resultant HCN(4-3)/HCO$^+$(4-3) and HCN(4-3)/CS(7-6) line ratios of each data point 
(hereafter we denote as $R_{\rm HCN/HCO^+}$ and $R_{\rm HCN/CS}$, respectively), 
are summarized in Table \ref{tbl1} with relevant information. 

We categorized the target galaxies into three classes of nuclear activities, 
namely {\it{AGN}}, {\it{buried-AGN}}, and {\it{SB}}, based on the following criteria. 
\begin{itemize}
\item[-] AGN: galaxies with clear broad Balmer lines (including polarized ones), 
or those with prominent hard X-ray ($>$ 2 keV) point sources with time variability. 
Therefore, galaxies with {\it{conventional}} AGN signatures naturally belong to this category. 
\item[-] buried-AGN: galaxies showing little (or no) AGN signatures at X-ray and optical wavelengths, 
but have been claimed to possess AGNs which are deeply embedded in dust along virtually all sight-lines. 
These galaxies are identified at infrared wavelength by detections of, 
e.g., continuum emission from a hot ($\ga$ 200 K) dust component, 
deep silicate absorption feature, and small equivalent width of polycyclic aromatic hydrocarbon (PAH) emission (e.g., \citealt{2007ApJS..171...72I}). 
\item[-] SB: galaxies with no AGN signature but host prominent starbursts at their nuclear regions. 
\end{itemize}
Brief descriptions of each galaxy relating the above criteria are presented in Appendix-\ref{app-A}. 
In this classification, we do not take into account the dominance 
of AGN and SB activities of each galaxy in molecular gas heating. 
In Table \ref{tbl1}, one can find some high resolution sample only exhibit lower limits in $R_{\rm HCN/CS}$ 
because of the non-detections ($<$ 3$\sigma$) of CS(7-6) emission line. 
However, we include them in our sample taking the rarity of 
extragalactic interferometric observations of submillimeter dense gas tracers into account.  

As for NGC 1068, which is the nearby best-studied type-2 Seyfert galaxy, we used ALMA band 7 data 
retrieved from the ALMA Science Archive\footnote{\url{http://almascience.nao.ac.jp/aq/}} (ID = 2011.0.00083.S). 
Although this data was already presented extensively in \citet{2014A&A...567A.125G} and \citet{2014A&A...570A..28V}, 
we re-analyzed the data to obtain high resolution values of the $R_{\rm HCN/HCO^+}$ and $R_{\rm HCN/CS}$, 
since the exact values of these ratios are not presented in \citet{2014A&A...567A.125G}, 
and the ratios in \citet{2014A&A...570A..28V} were averaged ones with a 100 pc aperture. 
We used MIRIAD (\citealt{1995ASPC...77..433S}) for this analysis. 
The synthesized beams and the rms noises in channel maps of the target emission lines were 
typically 0$''$.5 $\times$ 0$''$.4 (corresponds to 35 pc $\times$ 28 pc at the assumed distance of NGC 1068 = 14.4 Mpc) 
and 2.5 mJy beam$^{-1}$, respectively. 
The rms noises and measured fluxes of these lines are in good agreement with the published data. 
We assume the absolute flux uncertainty to be 15\%. 

At the end of this Section and before constructing an up-dated submm-HCN diagram, 
we mention the likely-limited applicability of our molecular diagnostics to some buried-AGNs. 
Recent high resolution observations of both 
vibrationally ground ($v$ = 0) and excited ($v$ = 1) HCN emission lines 
towards heavily obscured nuclei of ULIRGs revealed 
severe self- and/or continuum-absorption features at $v$ = 0 (\citealt{2015arXiv150406824A}). 
In the case of self-absorption, it is hard to extract physical/chemical information from line ratios. 
Thus, any kind of energy diagnostics employing 
such absorbed lines will have a limited power. 
So far, such self-absorption features in our target emission lines 
have been observed only in heavily obscured nuclei for the case of extragalactic objects 
(e.g., Arp 220W with the line-of-sight H$_2$ column density is $N_{\rm H_2}$ $>$ 10$^{25}$ cm$^{-2}$; \citealt{2015ApJ...800...70S}). 
Therefore, this can be a central issue in buried-AGNs with steep temperature gradient in gas, 
whose line-of-sight hydrogen column densities are extremely large 
($\ga$ 10$^{25-26}$ cm$^{-2}$; e.g., \citealt{2013ApJ...764...42S}). 
Contrary to these galaxies, we consider such self-absorption 
will not be a severe problem in Compton thin AGNs 
primarily because of the low optical depth of HCN(4-3) emission 
($\sim$ a few in NGC 1097 and NGC 7469, \citetalias{2013PASJ...65..100I}; \citealt{2015ApJ...811...39I}). 
This is in clear contrast to the extremely 
high optical depth ($\ga$ 100) in Arp 220W (\citealt{2015ApJ...800...70S}). 
Moreover, no such absorption feature was found even in the Compton thick AGN of NGC 1068 
(total hydrogen column density derived by X-ray observations $N_{\rm H}$ $\sim$ 10$^{25}$ cm$^{-2}$; \citealt{2012ApJ...748..130M,2015arXiv151103503M})
\footnote{
The multi-line analysis by \citet{2014A&A...570A..28V} suggested 
CO column density to velocity width ratio to be 3 $\times$ 10$^{17}$ cm$^{-2}$ (km s$^{-1}$)$^{-1}$ (see their Table 6). 
With the line width of $\sim$ 200 km s$^{-1}$ (\citealt{2014A&A...567A.125G}) and the assumption 
of CO fractional abundance of 10$^{-4}$ (e.g., \citealt{1987ApJ...315..621B}), 
this result corresponds to H$_2$ column density of $\sim$ 6 $\times$ 10$^{23}$ cm$^{-2}$. 
This value is $\sim$ one order of magnitude smaller than that of the X-ray derived 
total hydrogen column density (\citealt{2012ApJ...748..130M,2015arXiv151103503M}). 
However, we suggest this inconsistency would not be a problem 
because the averaged column density over the $\sim$ 100 pc beam employed by \citet{2014A&A...570A..28V} 
would show the lower limit of the nuclear (X-ray) obscuration 
operating at a much smaller scale (see also \citet{2004Natur.429...47J} for the expected size of the {\it{dusty torus}} of NGC 1068).
} 
observed at as high as 35 pc resolution (the data used in this work). 
Clearly, very extreme conditions are required to yield that feature. 

With carefully paying attention to these facts, 
we still keep using our buried-AGN samples in this work 
because we can not identify such absorption features 
in their spectrum at this moment (\citealt{2010ApJ...725L.228S,2013ApJ...764...42S,2014AJ....148....9I}). 
These galaxies are not used for a detailed quantitative discussion, 
but are used only to see an overall trend of line ratios. 
On the other hand, our discussion in the following is mostly based on 
the high resolution sample of AGNs and SB galaxies. 
Therefore, inclusion of our buried-AGN samples will not harm our conclusion. 
Of course, we admit it is plausible that these buried-AGNs would show 
absorption features when they are observed at higher spatial resolutions, 
but a quantitative assessment of this point is beyond the scope of this paper. 
Note that optically thinner emission and their ratios (e.g., H$^{13}$CN/H$^{13}$CO$^+$ ratio) seem to 
elucidate the nuclear physical/chemical conditions more straightforwardly in the case of obscured systems. 
We leave these caveats to future high resolution observations with ALMA.

\begin{table*}
\begin{center}
\caption{$R_{\rm HCN/HCO^+}$ and $R_{\rm HCN/CS}$ in AGNs, buried-AGNs, and SB galaxies \label{tbl1}}
\begin{tabular}{cccccccc}
\tableline\tableline
Object$^a$ & Distance$^b$ & Type$^c$ & Telescope$^d$ & Spatial resolution$^e$ & $R_{\rm HCN/HCO^+}$$^f$ & $R_{\rm HCN/CS}$$^g$ & Reference$^h$ \\
 & [Mpc] & & & [pc]& & & \\
\hline \hline
\multicolumn{8}{c}{High resolution ($<$ 500 pc) sample}\\ 
\tableline
NGC 1068 (AGN) & 14.4 & AGN & ALMA & 35 & 1.53$\pm$0.34 & 8.84$\pm$2.51 & (1) \\
NGC 1068 (E-knot) & 14.4 & AGN & ALMA & 35 & 2.84$\pm$0.60 & 8.04$\pm$1.72 & (1) \\
NGC 1068 (W-knot) & 14.4 & AGN & ALMA & 35 & 3.19$\pm$0.71 & 11.97$\pm$3.45 & (1) \\
NGC 1068 (CND-N) & 14.4 & AGN & ALMA & 35 & 3.14$\pm$0.72 & 12.79$\pm$4.48 & (1) \\
NGC 1068 (CND-S) & 14.4 & AGN & ALMA & 35 & 2.58$\pm$0.19 & $>$4.21 & (1) \\
NGC 1097 (AGN) & 14.5 & AGN & ALMA & 94 & 2.01$\pm$0.29 & $>$12.66 & (2) \\
NGC 7469 (AGN) & 70.8 & AGN & ALMA & 154 & 1.11$\pm$0.13 & 9.50$\pm$3.02 & (3) \\
M82 & 5.2 & SB & JCMT & 353 &  0.41$\pm$0.12 & 4.09$\pm$1.14 & (2) \\
NGC 253 & 3.0 & SB & JCMT, APEX & 262 & 1.03$\pm$0.22 & 3.40$\pm$0.73 & (2) \\
NGC 1097 (SB ring) & 14.5 & SB & ALMA & 94 & 0.82$\pm$0.17 & $>$1.38 & (2) \\
NGC 1614 & 69.1 & SB & ALMA & 468 & 0.24$\pm$0.06 & $>$3.54 & (4) \\
NGC 7469 (SB ring position-B) & 70.8 & SB & ALMA & 154 & 0.75$\pm$0.14 & $>$3.65 & (3) \\
NGC 7469 (SB ring position-C) & 70.8 & SB & ALMA & 154 & 0.45$\pm$0.08 & 2.50$\pm$1.16 & (3) \\
NGC 7469 (SB ring position-D) & 70.8 & SB & ALMA & 154 & 0.48$\pm$0.06 & 3.03$\pm$1.19 & (3) \\
N113 (LMC) & 0.05 & SB & ASTE & 5.3 & 0.21$\pm$0.06 & 1.30$\pm$0.58 & (5) \\
N159 (LMC) & 0.05 & SB & ASTE & 5.3 & 0.11$\pm$0.03 & 2.16$\pm$0.60 & (6) \\
\hline \hline
\multicolumn{8}{c}{Low resolution ($>$ 500 pc) sample}\\ 
\hline 
NGC 1068 (APEX) & 14.4 & AGN & APEX & 1257 & 1.85$\pm$0.42 & 5.00$\pm$1.47 & (7) \\
NGC 1365 & 16.9 & AGN & APEX & 1475 & 0.59$\pm$0.15 & 2.83$\pm$1.17 & (7) \\
NGC 4418 & 31.3 & buried-AGN & SMA & 790 & 1.64$\pm$0.25 & 1.98$\pm$0.30 & (2) \\
NGC 4945 & 5.2 & AGN & APEX & 454 & 0.78$\pm$0.17 & 3.21$\pm$0.70 & (7) \\
IRAS12127-1412 & 627.4 & buried-AGN & ALMA & 1667 & 1.58$\pm$0.58 & 4.32$\pm$2.17 & (8) \\
NGC 3256 & 37.4 & SB & APEX & 3265 & 0.39$\pm$0.11 & 1.38$\pm$0.51 & (7) \\
NGC 3628 & 7.7 & SB & APEX & 672 & 0.36$\pm$0.11 & 1.80$\pm$0.90 & (7) \\
NGC 7552 & 19.5 & SB & APEX & 1702 & 0.54$\pm$0.15 & 2.17$\pm$0.92 & (7) \\
IRAS13242-5713 & 42.0 & SB & APEX & 3667 & 1.23$\pm$0.29 & 2.39$\pm$0.66 & (7) \\
\tableline
\end{tabular}
\tablecomments{
$^{(a)}$For NGC 1068, we list the data obtained with both ALMA (archival data: ID=2011.0.00083.S) and APEX. 
The ratios of ALMA data are extracted at the positions of the {\it{AGN}}, {\it{E-knot}}, {\it{W-knot}}, {it{CND-N}}, and {it{CND-S}}. 
(see their coordinates in \citet{2014A&A...567A.125G} or \citet{2014A&A...570A..28V}). 
For NGC 1097 and NGC 7469, the ratios extracted both at the AGN position and the circumnuclear SB ring are listed. 
See their locations in \citetalias{2013PASJ...65..100I} and \citet{2015ApJ...811...39I}. 
Brief descriptions of each object are presented in Appendix-\ref{app-A}. 
$^{(b)}$We adopt distances determined by the Tully-Fisher relation (\citealt{1988ngc..book.....T}) 
for most cases since the local gravitational potentials can dominate the Hubble flow in the nearby universe. 
For NGC 4418, NGC 7469, IRAS12127-1412, NGC 1614, and IRAS13242-5713, 
distances are calculated based on their redshift recorded in NASA/IPAC Extragalactic Database (NED, \url{http://ned.ipac.caltech.edu}). 
We adopt $H_0$ = 70 km s$^{-1}$ Mpc$^{-1}$, $\Omega_{\rm M}$ = 0.27, and $\Omega_{\rm \Lambda}$ = 0.73 cosmology here. 
For N113 and N159, the Cepheid-based distance to the LMC (\citealt{2006ApJ...652.1133M}) is used. 
$^{(c)}$Type of a target galaxy (see also Section \ref{sec2} and Appendix-\ref{app-A}). 
$^{(d)}$The telescopes used for the observations. 
$^{(e)}$We separated the sample into two classes based on the spatial resolution of each observation. 
The resolution of 500 pc is employed here as the threshold, 
which is sufficient to fully encompass the typical size of CNDs of nearby AGNs. 
In the case of NGC 4945, we classified it into the low resolution sample 
despite the moderate spatial resolution to measure its line ratios (18$''$ $\sim$ 450 pc), 
because this galaxy hosts a relatively compact circumnuclear SB ring (2$''$.5 radius). 
For interferometric data, we list the geometrical mean of the FWHM of the major and minor axes of the synthesized beams. 
$^{(f)(g)}$Integrated intensity ratios of HCN(4-3)/HCO$^+$(4-3) and HCN(4-3)/CS(7-6) in the brightness temperature scale. 
The systematic errors are taken into account. 
We assume 15\% systematic error if it is not mentioned in the references. 
As for the $R_{\rm HCN/HCO^+}$ in NGC 4418, we mention that 0$''$.5 observations of \citet{2013ApJ...764...42S} found it to be $\sim$ 2. 
$^{(h)}$References for the $R_{\rm HCN/HCO^+}$ and $R_{\rm HCN/CS}$; 
(1) This work, but also see \citet{2014A&A...567A.125G} and \citet{2014A&A...570A..28V} for NGC 1068, 
(2) \citetalias{2013PASJ...65..100I} and references therein, 
(3) \citet{2015ApJ...811...39I}, 
(4) \citet{2013AJ....146...47I}, 
(5) \citet{2014A&A...572A..56P}, 
(6) \citet{2015arXiv151001246P}
(7) \citet{2014ApJ...784L..31Z}, 
(8) \citet{2014AJ....148....9I}. 
}
\end{center}
\end{table*}

\section{Up-dated Submillimeter-HCN diagram}\label{sec3}
\begin{figure*}
\epsscale{1}
\plotone{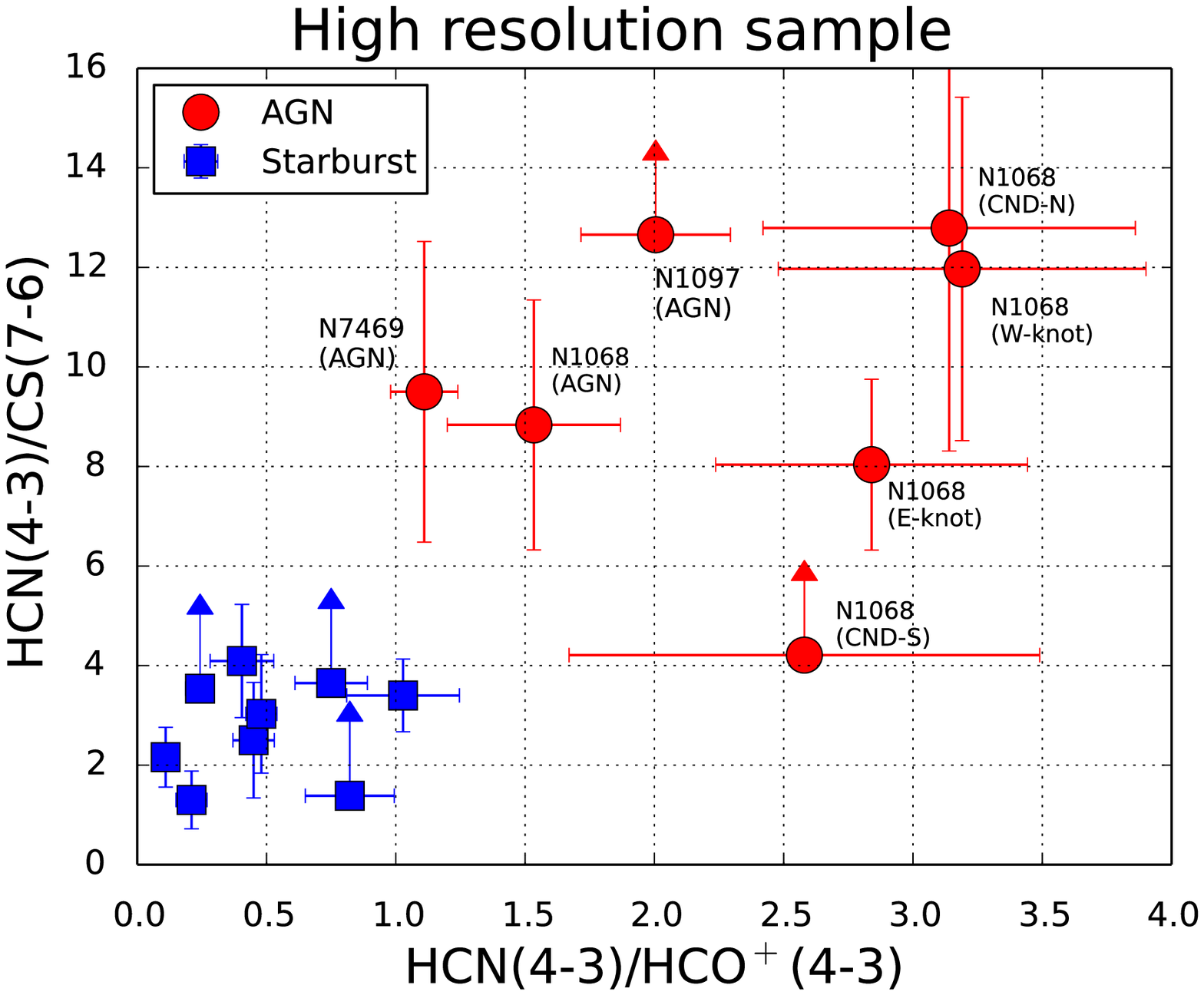}
\caption{Submillimeter-HCN diagram using HCN(4-3)/HCO$^+$(4-3) and HCN(4-3)/CS(7-6) integrated intensity ratios 
($R_{\rm HCN/HCO^+}$ and $R_{\rm HCN/CS}$ in the text, respectively) in the brightness temperature scale. 
Only the data obtained with high resolution observations (spatial resolution $<$ 500 pc, except for NGC 4945) are used. 
The red circles and the blue squares indicate AGNs and SB galaxies, respectively. 
The abbreviated names of AGNs are shown. 
Here, the term $``$AGN$"$ simply means that the galaxy hosts an AGN, 
regardless of its dominance in the total energy budget of the galaxy. 
See Table \ref{tbl1} for the details of the data. 
The systematic errors are also included here.}
\label{figure1}
\end{figure*}

Based on the data in Table \ref{tbl1}, 
we here up-date the submm-HCN diagram proposed by \citetalias{2013PASJ...65..100I}. 
We first show the result using only the high resolution sample 
(spatial resolution $<$ 500 pc) in Figure \ref{figure1}, 
to avoid strong contamination from the surrounding SB regions as much as possible. 
As a result, one can find a clear trend that AGNs exhibit higher $R_{\rm HCN/HCO^+}$ and/or $R_{\rm HCN/CS}$ than SB galaxies, 
which supports our previous claim in \citetalias{2013PASJ...65..100I}. 

\begin{figure*}
\epsscale{1}
\plotone{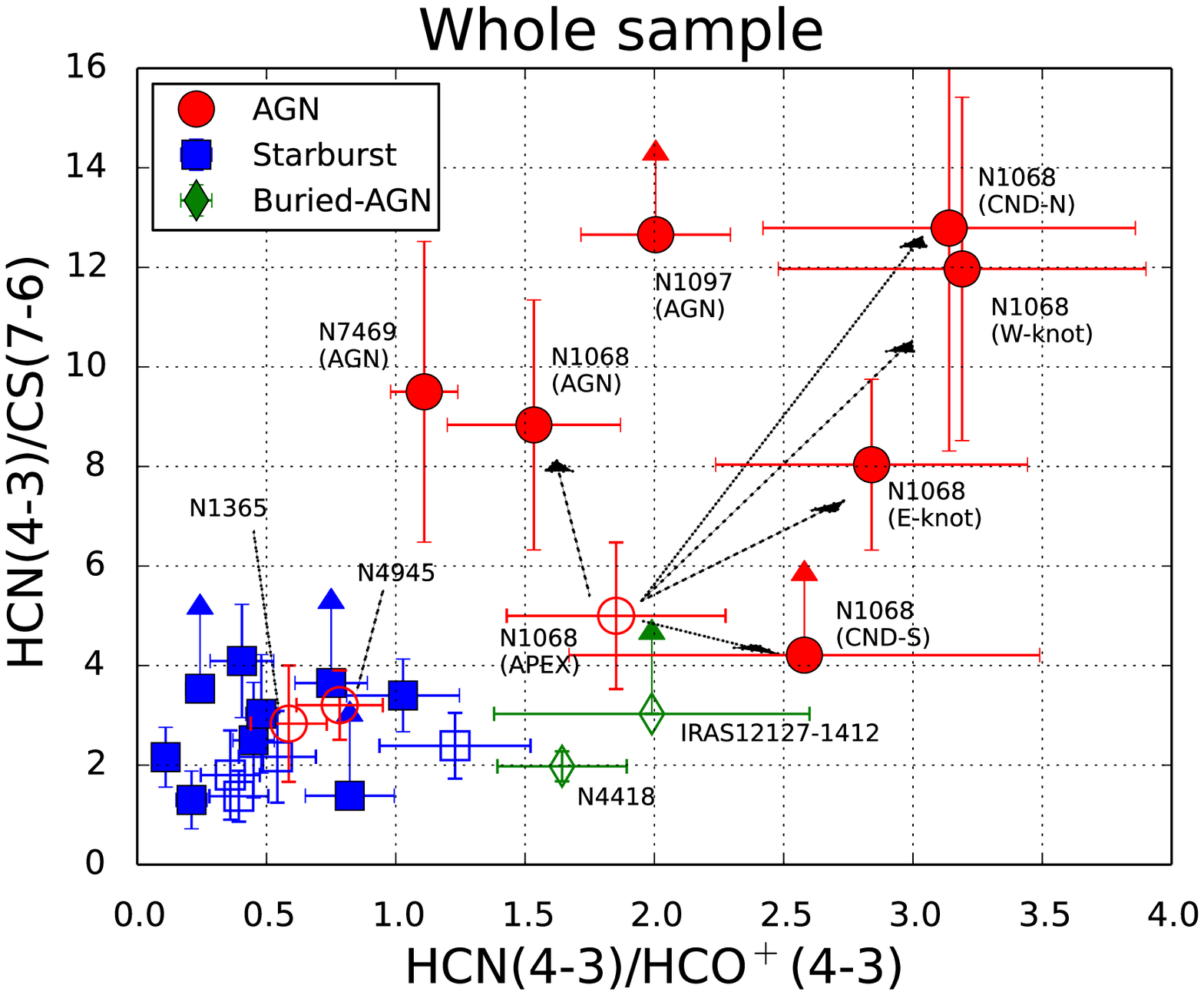}
\caption{
Same as Figure \ref{figure1}, 
but the whole sample including both the high resolution (spatial resolution $<$ 500 pc; filled symbol) 
and the low resolution (spatial resolution $>$ 500 pc; empty symbol) data are plotted. 
The red circles, green diamonds, and the blue squares indicate AGNs, buried-AGNs, and SB galaxies, respectively (see also Table \ref{tbl1}). 
The abbreviated names of AGNs and buried-AGNs are shown. 
See Table \ref{tbl1} for the details of the data. 
The systematic errors are included here.
}
\label{figure2}
\end{figure*}

The influence of spatial resolution on this line diagnostics is investigated in Figure \ref{figure2}, 
with superposing the low resolution sample (spatial resolution $>$ 500 pc). 
We found SB galaxies continue to show lower $R_{\rm HCN/HCO^+}$ and $R_{\rm HCN/CS}$ than most AGNs. 
On the other hand, NGC 1365 and NGC 4945 (both are AGNs) show 
line ratios fully comparable to SB galaxies, which is in contrast 
to the trend of the high resolution AGN sample. 
We suspect that in these two Seyfert galaxies, 
contamination from co-existing SB regions in line fluxes would 
be substantial when observed at the APEX 18$''$ beam. 
Indeed, both NGC 1365 and NGC 4945 host a prominent circumnuclear SB ring with a radius of 
5$''$--10$''$ (NGC 1365) and 2$''$.5 (NGC 4945) associating large amount of molecular gas 
(\citealt{2000A&A...357...24M,2005A&A...438..803G,2007ApJ...670..116C,2007ApJ...654..782S,2012MNRAS.425..311A}). 
Regarding the energetics, the equivalent widths of the 11.3 $\mu$m PAH feature are 432 nm 
(with 20$''$.4 $\times$ 15$''$.3 aperture; \citealt{2009ApJ...701..658W}) in NGC 1365 
and 358 nm (with 3$''$.7 slit; \citealt{2014ApJ...780...86E}) in NGC 4945, respectively. 
These widths are significantly larger than those of NGC 1068 (9 nm with 0$''$.36 slit, i.e., similar to the ALMA beam) 
and NGC 7469 (31 nm with 0$''$.75 slit, i.e., similar to the ALMA beam), for example (\citealt{2014ApJ...780...86E}). 
Moreover, the 25 $\mu$m-to-60 $\mu$m $IRAS$ colors in NGC 1365 and NGC 4945 are 0.14 and 0.04, respectively. 
These equivalent widths and IR-colors are clearly categorized in the SB regime (\citealt{2009ApJ...701..658W}). 
Therefore, the low line ratios in NGC 1365 and NGC 4945 compared to the high resolution AGN samples 
would highlight the importance of high spatial resolution (likely to be $\la$ 50--100 pc scale in these cases) 
to robustly identify low luminosity AGNs accompanying prominent circumnuclear SB based on this diagram. 
This would reflect the limited spatial extent of energetic influence 
of AGNs such as XDRs (\citealt{2010A&A...513A...7S,2015ApJ...811...39I}). 

On the other hand, once we achieve the high resolution, 
we should carefully treat the spatially resolved measurements of the line ratios 
because they would reflect very local physics and/or underlying chemistry 
even within a single AGN environment (\citealt{2014A&A...570A..28V}). 
This is clearly manifested by NGC 1068 (Figure \ref{figure2}); 
both $R_{\rm HCN/HCO^+}$ and $R_{\rm HCN/CS}$ measured with ALMA (0$''$.5 beam) 
at the different positions within the CND are different from those with APEX (18$''$ beam). 
Contrary to this case, spatial resolution seems not to play an important role 
for the ratios of SB galaxies because both the high and the low resolution samples 
exhibit comparable line ratios as already mentioned (see also Table \ref{tbl1}). 
This could be reconciled if a SB region has a more extended nature 
(= ensemble of massive star forming regions) than a compact CND around an AGN. 

Considering the above, we suggest from Figures \ref{figure1} and \ref{figure2} 
that (1) galaxies energetically dominated by AGNs show enhanced $R_{\rm HCN/HCO^+}$ and/or $R_{\rm HCN/CS}$, 
and (2) those by SB show lower values in both ratios than AGNs. 
We also point out that the buried-AGNs of our sample 
tend to exhibit relatively high $R_{\rm HCN/HCO^+}$ ($\ga$ 1.5) 
but rather low $R_{\rm HCN/CS}$ ($\sim$ a few) which is comparable to SB galaxies. 
These buried-AGNs belong to our low resolution sample. 
Thus, one concern is that the line ratios will change when observed at a higher resolution. 
However, at least for NGC 4418, we suppose that such a situation would be unlikely 
because $\sim$ 100 pc scale observations of HCN(4-3) and HCO$^+$(4-3) revealed 
that the dense molecular gas is well confined 
in the central $\sim$ 100 pc region (\citealt{2013ApJ...764...42S,2015A&A...582A..91C}). 
Therefore, the location of this galaxy will more or less hold 
in Figure \ref{figure2} even when observed at $\sim$ 100 pc resolution, 
although we need to increase the high resolution sample 
of buried-AGNs to examine their overall trend in this diagram. 
In the following Sections, we will investigate possible causes for the HCN-enhancement in AGNs, 
from the perspectives of both {\it{line excitation}} and {\it{abundance (ISM chemistry)}}. 
The line ratios of the high resolution samples shown in Figure \ref{figure1} 
should be the reference for the discussion in the following 
as those of the low resolution samples (especially AGNs) 
are highly likely to be contaminated by various other components. 
We should note that the observed line ratios are the integrated ones over 
not only some areas but also the line-of-sight columns, 
thus, all physical and chemical gradients are integrated. 

\section{Non-LTE Excitation Analysis}\label{sec4}
In order to investigate the physical origin of the HCN-enhancement, 
we ran non-LTE radiative transfer models with the RADEX code (\citealt{2007A&A...468..627V}). 
The RADEX uses an escape probability approximation to treat optical depth effects and solves statistical equilibrium 
in a homogeneous (i.e., single temperature and density), one-phase medium. 
Therefore, all HCN(4-3), HCO$^+$(4-3), and CS(7-6) lines are emitted from the same volume in our models. 
This assumption would not be very crude considering the relatively narrow range of the $n_{\rm cr}$ (Table \ref{tbl2}) 
and similar velocity profiles of these lines (e.g., \citealt{2014ApJ...784L..31Z}). 
As for the cloud geometry, we assumed a spherical one. 
Other relevant excitation parameters of the target lines are summarized in Table \ref{tbl2}. 
We hereafter express line strengths in the brightness temperature scale. 
Note that we do not intend to mimic an environment of a specific galaxy here. 
Moreover, one line ratio can be reproduced by various combinations of parameters. 
Hence the model described below is the result of an educated guess of the parameters, 
which should be further investigated with future observations. 

\begin{table*}
\begin{center}
\caption{The excitation parameters of HCN(4-3), HCO$^+$(4-3), and CS(7-6) line emissions \label{tbl2}}
\begin{tabular}{ccccccccc}
\tableline\tableline
Line name$^a$ & $\nu_{\rm rest}$ [GHz]$^b$ & $\mu$ [Debye]$^c$ & $E_u$/$k_{\rm B}$ [K]$^d$ & $\Delta E_{ul}$ [K]$^e$ 
& $A_{ul}$ [s$^{-1}$]$^f$ & \multicolumn{3}{c}{$n_{\rm cr,thin}$ [cm$^{-3}$]$^g$} \\
 & & & & & & $T_{\rm kin}$ = 50 K & $T_{\rm kin}$ = 100 K & $T_{\rm kin}$ = 200K \\
\tableline
HCN($J$=4-3) & 354.505 & 2.99 & 42.5 & 17.0 & 2.054 $\times$ 10$^{-3}$ & 1.4 $\times$ 10$^7$ & 9.1 $\times$ 10$^6$ & 6.1 $\times$ 10$^6$ \\
HCO$^+$($J$=4-3) & 356.734 & 3.93 & 42.8 & 17.1 & 3.627 $\times$ 10$^{-3}$ & 2.6 $\times$ 10$^6$ & 2.0 $\times$ 10$^6$ & 1.6 $\times$ 10$^6$ \\
CS($J$=7-6) & 342.883 & 1.96 & 65.8 & 16.4 & 8.395 $\times$ 10$^{-4}$ & 3.4 $\times$ 10$^6$ & 2.6 $\times$ 10$^6$ & 2.2 $\times$ 10$^6$ \\
\tableline
\end{tabular}
\tablecomments{
$^{(a)}$Full name of the line. 
$^{(b)(c)(d)(e)(f)}$The rest frequency, dipole moment, upper level energy, energy gap between the upper and lower levels, 
and Einstein A-coefficient of the transition, respectively. 
These values are extracted from the Leiden Atomic and Molecular Database ({\it{LAMDA}}: \citealt{2005A&A...432..369S}). 
$^{(g)}$Critical density of the line in the optically thin limit without any background radiation, 
which is calculated for the kinetic temperature ($T_{\rm kin}$) of 50, 100, and 200 K, using 
$n_{\rm cr,thin}$ = $A_{jk}$/[$\Sigma_{i<j}$$\gamma_{ji}$ + $\Sigma_{i>j}$($g_i/g_j$)$\gamma_{ij}$exp(-($E_i$-$E_j$)/$T_{\rm kin}$)] 
for the $j$ $\rightarrow$ $k$ transition. 
Here, $\gamma_{jk}$ indicates the collision rate of the $j$ $\rightarrow$ $k$ transition. 
We adopt only H$_2$ for the collision partner and include collisional de-excitation as well from $J_{\rm upper}$ up to 25. 
Values for the collision rate $\gamma_{\rm ul}$ are also extracted from {\it{LAMDA}}. 
}
\end{center}
\end{table*}

\subsection{Model description}\label{sec4.1}
In our RADEX simulation, we investigated how the following parameters affect the line ratios of our interest. 
\begin{itemize}
\item Kinetic temperature ($T_{\rm kin}$) of the molecular gas: 
this affects the rate of the collisional excitation with the target molecules. 
The cases of 50, 100, and 200 K are investigated. 
This range mostly covers the $T_{\rm kin}$ suggested for nearby AGNs and SB galaxies 
(e.g., \citealt{2003A&A...403..561M,2008ApJ...677..262K,2012A&A...537A.133D,2014A&A...570A..28V}). 
\item Molecular gas density ($n_{\rm H_2}$): this also affects the rate of collisional excitation. 
Two cases of $n_{\rm H_2}$ = 10$^5$ and 5 $\times$ 10$^6$ cm$^{-3}$ will be examined. 
These values are typical ones suggested in nuclear regions of galaxies 
(e.g., \citetalias{2013PASJ...65..100I}; \citealt{2014A&A...570A..28V}), 
which is also supported by the commonly sub-thermal excitation of our target lines 
(e.g., \citetalias{2013PASJ...65..100I}; \citealt{2007ApJ...666..156K,2014A&A...570A..28V}). 
\item Ratios of molecular fractional abundances with respect to H$_2$ ($X_{\rm mol}$, mol = HCN, HCO$^+$, and CS): 
we will show two cases for simplicity, where $X_{\rm HCN}/X_{\rm HCO^+}$ (or $X_{\rm HCN}/X_{\rm CS}$) = 1 and 10. 
\item Background radiation temperature ($T_{\rm bg}$): 
molecular rotational levels can be radiatively excited through absorbing photons. 
In this perspective, it is highly likely that the background radiation is stronger around AGNs than in SB environments. 
In fact, the dust temperature at FIR to submillimeter (i.e., Rayleigh-Jeans regime) is 
as high as 46 K in the central $\sim$ 400 pc region of NGC 1068 (AGN; \citealt{2014A&A...567A.125G}), 
but 29 K in the central 1.2 kpc region of NGC 253 (SB; \citealt{2008A&A...490...77W}), for example. 
The cases of 2.73 (Cosmic Microwave Background = CMB), 5, 10, 20, 30, 40, 50, and 60 K are studied. 
We simply adopt the black body approximation to calculate the background radiation field. 
\item Optical depth of the line emission ($\tau$): 
models with different $N_{\rm mol}$/$dV$ 
(or equivalently a volume density of the target molecule to a velocity gradient ratio) 
are employed to test this effect. 
Here, $N_{\rm mol}$ and $dV$ are a line-of-sight column density and a line velocity width, respectively. 
$N_{\rm HCO^+}$/$dV$ (or $N_{\rm CS}$/$dV$) = 5.0 $\times$ 10$^{12}$, 5.0 $\times$ 10$^{13}$, 
5.0 $\times$ 10$^{14}$, and 5.0 $\times$ 10$^{15}$ cm$^{-2}$ (km s$^{-1}$)$^{-1}$ are studied. 
The $N_{\rm HCN}$/$dV$ is equated to the above $N_{\rm HCO^+}$/$dV$ (or $N_{\rm CS}$/$dV$), or enhanced by 10-times, 
since $X_{\rm HCN}$/$X_{\rm HCO^+}$ (or $X_{\rm HCN}$/$X_{\rm CS}$) = 1 and 10 as already mentioned. 
\end{itemize}

Under these conditions, we ran the RADEX for each set of 
($T_{\rm kin}$, $n_{\rm H_2}$, $X_{\rm HCN}$/$X_{\rm HCO^+}$, $T_{\rm bg}$, $N_{\rm HCO^+}$/$dV$) 
or ($T_{\rm kin}$, $n_{\rm H_2}$, $X_{\rm HCN}$/$X_{\rm CS}$, $T_{\rm bg}$, $N_{\rm CS}$/$dV$) 
and took line ratios of $R_{\rm HCN/HCO^+}$ and $R_{\rm HCN/CS}$. 
Transitions between vibrational levels through IR-pumping are not included in the models 
since there is no detection of vibrationally excited HCN emission 
in our sample galaxies except for NGC 4418 (\citealt{2010ApJ...725L.228S}), 
although it can be a bit inappropriate treatment in some cases (Section \ref{sec5}). 
Regarding the gas density, we here mention that \citet{2008ApJ...677..262K} 
suggested lower $n_{\rm H_2}$ in AGNs than in SB galaxies, 
but their measurements were based on single dish observations of, e.g., HCN(3-2) to HCN(1-0) line ratio. 
On the other hand, we need spatially resolved measurements of line ratios 
to accurately assess $n_{\rm H_2}$ or line excitation (\citealt{2014A&A...570A..28V}). 
One may expect that an increase in $n_{\rm H_2}$ naturally leads to higher line ratios 
since $n_{\rm cr}$ of HCN(4-3) is the highest among the target lines. 
However, this is actually not so straightforward as shown later. 

In the modeling below, we first fix $n_{\rm H_2}$ to be 10$^5$ cm$^{-3}$ 
and investigate the dependence of the line ratios on the other parameters. 
The case of $n_{\rm H_2}$ = 5 $\times$ 10$^6$ cm$^{-3}$ will be shown subsequently. 
Note that our modelling is fundamentally different from the LTE modelings by, 
e.g., \citetalias{2013PASJ...65..100I}, \citet{2014A&A...570A..28V}, and \citet{2015A&A...573A.116M}, 
in the sense that non-LTE processes are treated here. 
Moreover, we also examine the dependence of the line ratios on $T_{\rm bg}$, 
which has not been studied in the non-LTE modelings by, e.g., \citet{2011ApJ...736...37K} and \citet{2014A&A...570A..28V}. 
Hence, in addition to the previous key works, we expect 
that our analysis will provide some insight on 
the underlying physical and chemical conditions in the centers of galaxies.

\subsection{Molecular line excitation with photon trapping}\label{sec4.2}
\begin{figure*}
\epsscale{1.2}
\plotone{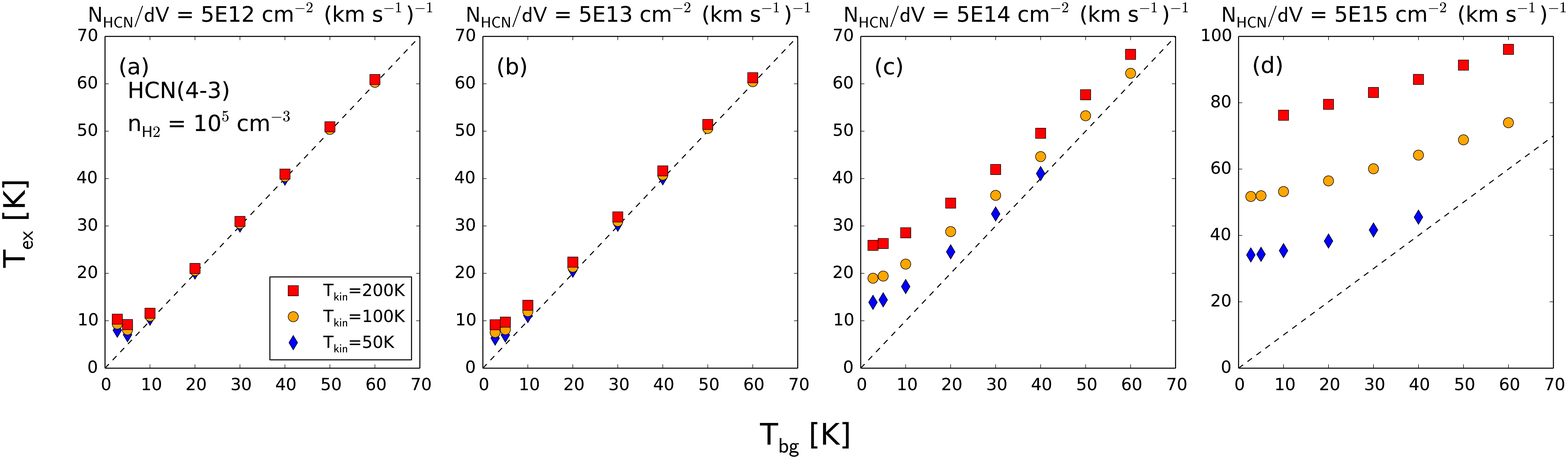}
\caption{Excitation temperature ($T_{\rm ex}$) of HCN(4-3) as a function of background temperature ($T_{\rm bg}$). 
The red, orange, blue symbols indicate the models with gas kinetic temperature ($T_{\rm kin}$) of 200, 100, and 50 K, respectively. 
We here fixed gas volume density as $n_{\rm H_2}$ = 10$^5$ cm$^{-3}$. 
Four cases of the line-of-sight column density to velocity width ratio ($N_{\rm HCN}$/$dV$) of 
(a) 5 $\times$ 10$^{12}$, (b) 5 $\times$ 10$^{13}$, (c) 5 $\times$ 10$^{14}$, and (d) 5 $\times$ 10$^{15}$ cm$^{-2}$ (km s$^{-1}$)$^{-1}$ are shown here. 
Note that the scale of the $y$-axis in the panel-(d) is different from the others. 
The dashed line in each panel indicates the $T_{\rm ex}$ = $T_{\rm bg}$. 
One can find that $T_{\rm ex}$ approaches to $T_{\rm bg}$ when $T_{\rm bg}$ $\ga$ 10 K in the panels-(a) and (b). 
In the panel-(d), optical depth is so large (see also Figure \ref{figure4}) 
that we can expect $T_{\rm ex}$ becomes independent of $T_{\rm bg}$, and approaches to $T_{\rm kin}$ 
due to an enhanced photon trapping effect. 
Note that HCN(5-4) (not $J$ = 4-3) line shows a maser feature 
($T_{\rm ex}$ $<$ 0) at $T_{\rm bg}$ = 2.73 K in the panel-(d), 
and HCN(4-3) is also a maser at $T_{\rm bg}$ = 5 K in the same panel, when $T_{\rm kin}$ = 200 K. 
These two extreme cases are excluded from the plot. 
}
\label{figure3}
\end{figure*}

\begin{figure*}
\epsscale{1.2}
\plotone{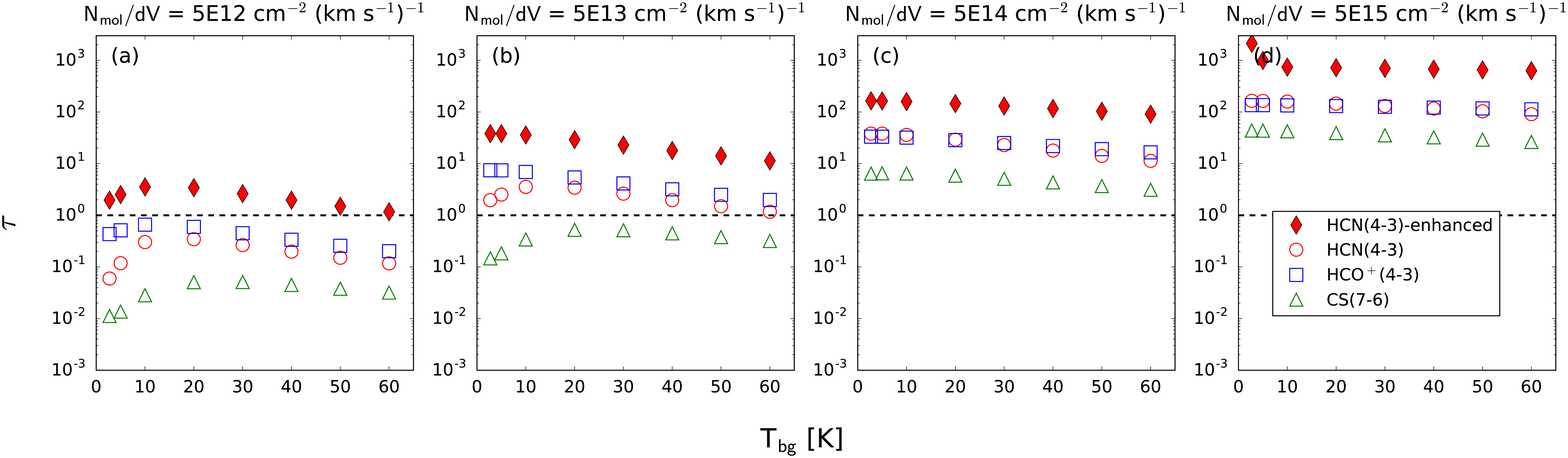}
\caption{
Line optical depths of HCN(4-3), HCO$^+$(4-3), and CS(7-6), as a function of $T_{\rm bg}$ calculated with the RADEX. 
The H$_2$ gas density and the kinetic temperature were fixed to $n_{\rm H_2}$ = 10$^5$ cm$^{-3}$ and $T_{\rm kin}$ = 100 K, respectively. 
The cases where the molecular column density to the velocity width ratio ($N_{\rm mol}$/$dV$, mol = HCN, HCO$^+$, and CS) of 
(a) 5.0 $\times$ 10$^{12}$, (b) 5.0 $\times$ 10$^{13}$, (c) 5.0 $\times$ 10$^{14}$, 
and (d) 5.0 $\times$ 10$^{15}$ cm$^{-2}$ (km s$^{-1}$)$^{-1}$ are presented. 
In each panel, we also show the optical depth of HCN(4-3) calculated after 
enhancing the abundance of HCN by 10 times (filled diamond) 
for easier comparison with the subsequent figures. 
The dashed line indicates $\tau$ = 1.0 for an eye-guide. 
}
\label{figure4}
\end{figure*}

Excitation states of the target lines under the fixed gas density of 
$n_{\rm H_2}$ = 10$^5$ cm$^{-3}$ are described here. 
HCN(4-3) is mainly used as a representative case, 
but the same argument can hold for HCO$^+$(4-3) and CS(7-6) as well. 
Figure \ref{figure3} shows excitation temperature ($T_{\rm ex}$) of HCN(4-3) as a function of $T_{\rm bg}$. 
When the line emission is optically thin or moderately thick (panels-(a) and (b) in Figure \ref{figure3}), 
one can find that the excitation is dominated by radiative processes, 
since $T_{\rm ex}$ is very close to $T_{\rm bg}$ especially at $T_{\rm bg}$ $\ga$ 10 K. 
Note that $\Delta$$E_{\rm ul}$ of HCN(4-3) is 17.0 K (Table \ref{tbl2}). 
The excitation states of HCO$^+$(4-3) and CS(7-6) are presented in Appendix-\ref{app-B}. 
The optical depths of the target lines calculated by the RADEX are shown in Figure \ref{figure4} as a function of $T_{\rm bg}$. 

The excitation state shown in Figure \ref{figure3} can be better understood 
through a two-level (e.g., HCN $J$ = 3 and 4) analytic treatment. 
Although we conducted full-level statistical equilibrium calculation with the RADEX here, 
this analytic treatment provides us fruitful insights on 
what is influencing the molecular excitation. 
Similar approach can be found in, e.g., \citet{2015ApJ...800...70S} and \citet{1974ApJ...187L..67S}, 
but we include background radiation into the analysis as well, which has been usually omitted. 

\begin{figure}
\epsscale{1}
\plotone{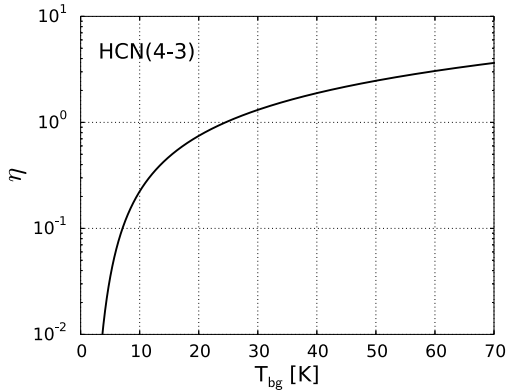}
\caption{
Dependence of $\eta$ $\equiv$ 1/(exp($\Delta E_{ul}$/$T_{\rm bg}$)-1) of HCN(4-3) on $T_{\rm bg}$. 
The value of $\eta$ is almost identical among HCN(4-3), HCO$^+$(4-3), and CS(7-6), 
reflecting their similar energy gaps between the upper and the lower levels (Table \ref{tbl2}). 
}
\label{figure5}
\end{figure}

The ratio of the upper to lower level molecular population with the energy gap of $\Delta E_{ul}$ can be written as, 
\begin{eqnarray}\label{eq1}
\frac{n_u}{n_l} = \frac{B_{lu} J_\nu + C_{lu}}{A_{ul} + B_{ul}J_\nu + C_{ul}} 
&=& \frac{g_u}{g_l} \frac{\displaystyle \eta A_{ul}\beta_\nu + C_{ul} \exp \left(-\frac{\Delta E_{ul}}{T_{\rm kin}}\right)}{\displaystyle (1 + \eta )A_{ul}\beta_\nu + C_{ul}} \nonumber \\
&=& \frac{g_u}{g_l} \exp \left(\displaystyle -\frac{\Delta E_{ul}}{T_{\rm ex}} \right),
\end{eqnarray}
with 
\begin{eqnarray}\label{eq2}
\eta \equiv \frac{1}{\exp \left(\displaystyle \frac{\Delta E_{ul}}{T_{\rm bg}}\right)-1}. 
\end{eqnarray}
Here, the $A_{ul}$, $B_{ul}$, and $C_{ul}$ indicate the Einstein coefficients of spontaneous decay, stimulated emission, 
and collisional de-excitation, respectively. 
The upwards and downwards collision rates are related assuming a detailed balance 
in the thermodynamic equilibrium state. 
The $g_u$ and $g_l$ are statistical weights of the upper and lower levels. 
The frequency of the line is represented as $\nu$. 
The internal radiation $J_\nu$ is 
\begin{equation}\label{eq3}
J_\nu = ( 1 - \beta_\nu ) B_\nu (T_{\rm ex}) + \beta_\nu B_\nu (T_{\rm bg}),
\end{equation}
where $B_\nu (T)$ is the Planck function at temperature $T$. 
Although we should include various mechanisms as the source of the background radiation 
and solve their radiative transfer individually to achieve the local spectral energy distribution, 
we represent them by a single Planck function throughout this paper for simplicity. 
The photon escape probability from the model cloud is denotes as $\beta_\nu$, 
which is a function of the line optical depth $\tau$ as 
\begin{equation}\label{eq4}
\beta_\nu = \frac{1.5}{\tau} \left[ 1 - \frac{2}{\tau^2} + \left(\displaystyle \frac{2}{\tau} + \frac{2}{\tau^2}\right) e^{-\tau} \right],
\end{equation}
for a spherical cloud (e.g., \citealt{2006agna.book.....O}). 
The value of $\beta_{\nu}$ also depends on the assumed geometry. 
In Equation (\ref{eq1}), $A_{ul}$ is reduced by $\beta_\nu$, 
which means an {\it{effective}} $n_{\rm cr}$ of the line ($n_{\rm cr,eff}$) 
is lower than the $n_{\rm cr}$ at the optically thin limit ($n_{\rm cr,thin}$; Table \ref{tbl2}) due to the {\it{photon trapping effect}}, 
i.e., $n_{\rm cr,eff}$ = $\beta_\nu \times n_{\rm cr,thin}$. 
The background radiation field is included in Equation (\ref{eq1}) as $\eta$. 
We show the $\eta$ of HCN(4-3) in Figure \ref{figure5}, 
which is almost identical to those of HCO$^+$(4-3) and CS(7-6). 

By introducing the $n_{\rm cr,eff}$, Equation (\ref{eq1}) is reduced to 
\begin{eqnarray}\label{eq5}
\exp \left(-\frac{\Delta E_{ul}}{T_{\rm ex}} \right) 
&= \frac{\displaystyle \eta + \left( \frac{n_{\rm H_2}}{n_{\rm cr,eff}} \right) \exp \left( -\frac{\Delta E_{ul}}{T_{\rm kin}}\right)}
{\displaystyle (1+ \eta ) +  \left( \frac{n_{\rm H_2}}{n_{\rm cr,eff}} \right) } \nonumber \\
&\sim \frac{\displaystyle \eta + \left( \frac{n_{\rm H_2}}{n_{\rm cr,eff}} \right)}
{\displaystyle (1+ \eta ) +  \left( \frac{n_{\rm H_2}}{n_{\rm cr,eff}} \right) }.
\end{eqnarray} 
Here exp(-$\Delta E_{ul}/T_{\rm kin}$) commonly takes $\sim$ 1 
for HCN(4-3), HCO$^+$(4-3), and CS(7-6) at any $T_{\rm kin}$ in our models. 
With this formula, the molecular excitation in the two extreme cases of $\beta_\nu$, 
i.e., $\beta_\nu \rightarrow 1$ (optically thin limit) and $\beta_\nu \rightarrow 0$ (optically thick limit), 
are discussed in the following. 

(i) optically thin limit---in this limit, we expect 
$n_{\rm H_2}/n_{\rm cr,eff}$ = $n_{\rm H_2}/n_{\rm cr,thin}$ $\sim$ 10$^{-2}$ 
for HCN(4-3), for example (Table \ref{tbl2}). 
Taking Figure \ref{figure5} into account as well, 
one can find both collisional and radiative excitation can contribute to the molecular excitation 
especially when $T_{\rm bg}$ $\la$ 5 K. 
In this range, $\eta$ $\la$ $n_{\rm H_2}/n_{\rm cr,eff}$. 
On the other hand, the right-hand-side of Equation (\ref{eq5}) reduces to $\exp (-\Delta E_{ul}/T_{\rm bg})$, 
i.e., molecules are radiatively excited. 
This trend stands out especially when $T_{\rm bg}$ $\ga$ 10 K, 
which can be clearly seen in Figure \ref{figure3}(a). 
Therefore, as a general manner, we suggest that radiative excitation 
should be considered seriously when we treat 
optically thin to moderately thick line emissions from AGNs, 
where high $T_{\rm bg}$ is likely expectable. 

(ii) optically thick limit---in this limit, by substituting 
$\beta_\nu$ $\rightarrow$ 0 (or $n_{\rm cr,eff}$ $\rightarrow$ 0) to Equation (\ref{eq5}), 
one can find that $T_{\rm ex}$ is now identical to $T_{\rm kin}$ and thus independent of $T_{\rm bg}$. 
Indeed, $T_{\rm ex}$ is less dependent on $T_{\rm bg}$ and also differs a lot according to $T_{\rm kin}$ (panels-(c) and (d) in Figure \ref{figure3}), 
as HCN(4-3) becomes optically thicker (see also Figure \ref{figure4}). 
We note that the output parameters have a limited meaning at high optical depth 
such as $\tau$ $\ga$ 100 in the RADEX code, since the change of optical depth over 
the line profile is not taken into account (\citealt{2007A&A...468..627V}).

\subsection{The $R_{\rm HCN/HCO^+}$ and $R_{\rm HCN/CS}$ under non-LTE}\label{sec4.3}
\begin{figure*}
\epsscale{1.2}
\plotone{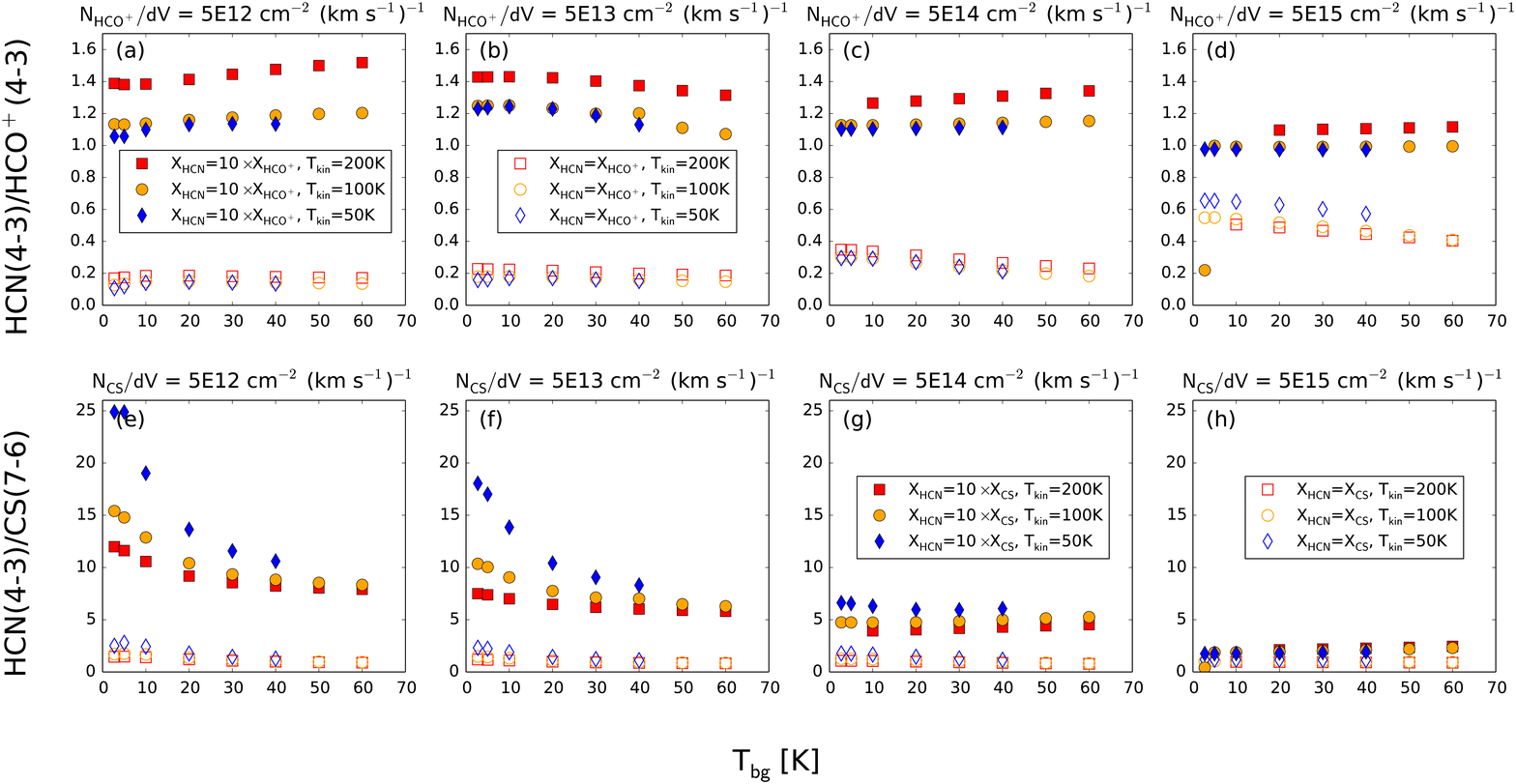}
\caption{
The $R_{\rm HCN/HCO^+}$ (top panels) and the $R_{\rm HCN/CS}$ (bottom panels) as a function of $T_{\rm bg}$, 
calculated by our RADEX models. 
The ratio of line-of-sight molecular column density to velocity width ($N_{\rm mol}$/$dV$) varies from 
5 $\times$ 10$^{12}$ to 5 $\times$ 10$^{15}$ cm$^{-2}$ (km s$^{-1}$)$^{-1}$ 
for the cases of mol = HCO$^+$ and CS, respectively. 
As a result, the panels (a) and (e) display the case where each emission is optically thin, 
whereas the panels (d) and (h) show heavily optically thick cases. 
The opacities of these lines are shown in Figure \ref{figure4} for the case of $T_{\rm kin}$ = 100 K. 
The filled and empty symbols indicate $X_{\rm HCN}/X_{\rm HCO^+}$ (or $X_{\rm HCN}/X_{\rm CS}$) = 10 or 1, respectively. 
The colors indicate the gas kinetic temperature ($T_{\rm kin}$) as blue = 50 K, orange = 100 K, and red = 200 K. 
}
\label{figure6}
\end{figure*}

We then calculated $R_{\rm HCN/HCO^+}$ and $R_{\rm HCN/CS}$ 
based on our non-LTE modelings as a function of $T_{\rm bg}$. 
Several cases with different $T_{\rm kin}$, $N_{\rm mol}$/$dV$ (or $\tau$), 
and molecular abundance ratios, are shown in Figure \ref{figure6}. 
Here, we discuss a dependence of each line ratio on the parameters in our models. 
We guide readers to \citet{2015A&A...573A.116M} for these line ratios calculated under the LTE condition. 

\begin{figure*}
\epsscale{1}
\plotone{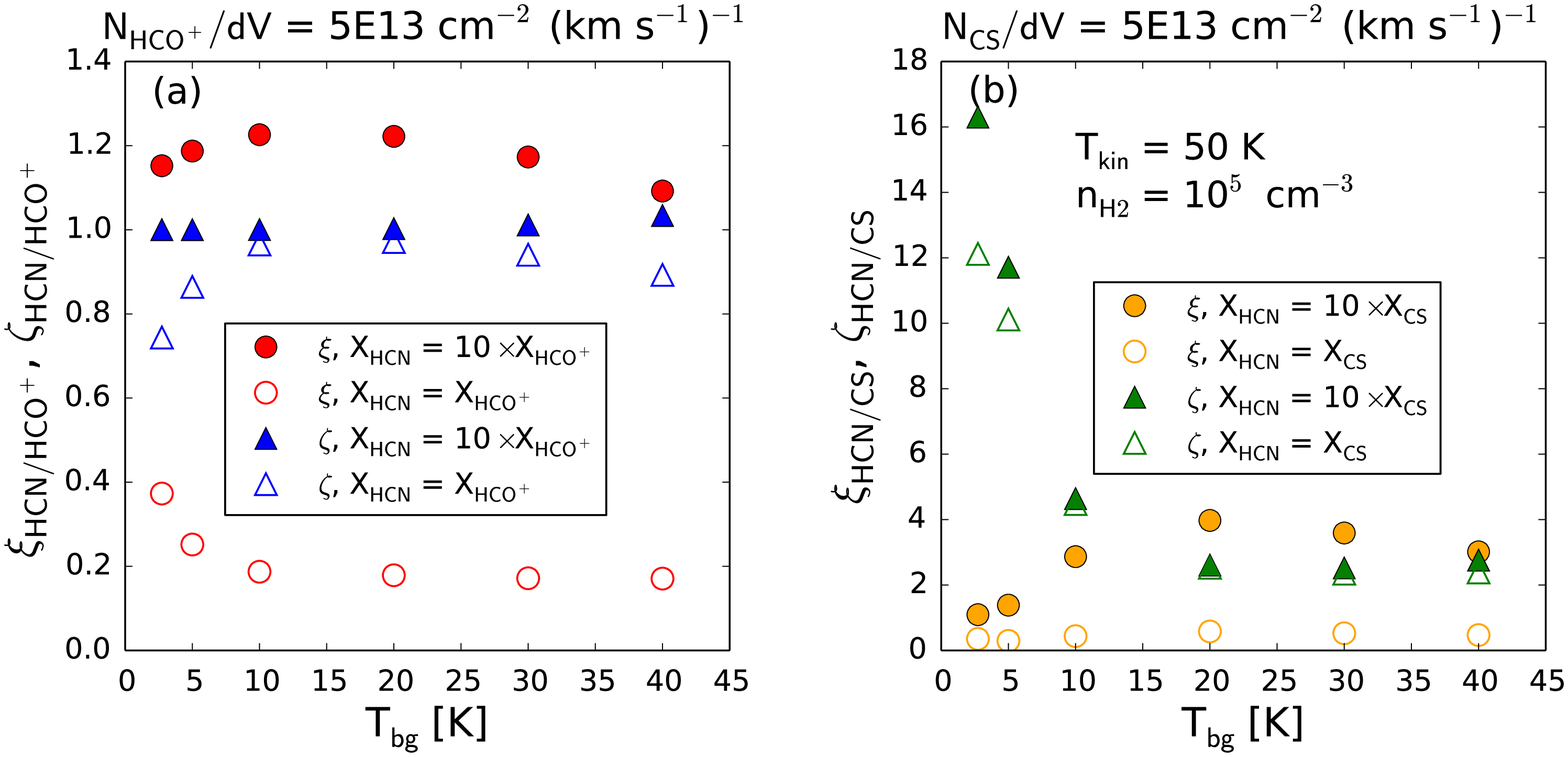}
\caption{
Dependence of (a) $\xi_{\rm HCN/HCO^+}$ and $\zeta_{\rm HCN/HCO^+}$, 
and (b) $\xi_{\rm HCN/CS}$ and $\zeta_{\rm HCN/CS}$ on $T_{\rm bg}$. 
For the definition of $\xi$ and $\zeta$, see Equations (\ref{eq6}) and (\ref{eq9}) in the text. 
The case of ($T_{\rm kin}$, $n_{\rm H_2}$) = (50 K, 10$^5$ cm$^{-3}$) 
with $N_{\rm mol}/dV$ = 5 $\times$ 10$^{13}$ cm$^{-2}$ (km s$^{-1}$)$^{-1}$ 
are shown (mol = HCO$^+$ or CS) as the representative ones. 
The colors indicate (a) red = $\xi_{\rm HCN/HCO^+}$, blue = $\zeta_{\rm HCN/HCO^+}$, 
and (b) orange = $\xi_{\rm HCN/CS}$, green = $\zeta_{\rm HCN/CS}$, respectively. 
The filled and empty symbols denote $X_{\rm HCN}/X_{\rm HCO^+}$ (or $X_{\rm HCN}/X_{\rm CS}$) = 10 and 1, respectively. 
}
\label{figure7}
\end{figure*}

(i) $R_{\rm HCN/HCO^+}$---
one will find in Figure \ref{figure6}(a)-(d) that this ratio is not so sensitive to $T_{\rm bg}$, 
which is close to $T_{\rm ex}$ when the excitation is dominated by the radiative processes (see also Figure \ref{figure3}). 
To further examine this trend, we rewrite $R_{\rm HCN/HCO^+}$ analytically as 
\begin{equation}\label{eq6}
\frac{T_{\rm ex,HCN(4-3)} - T_{\rm bg}}{T_{\rm ex,HCO^+(4-3)} - T_{\rm bg}} \cdot \frac{1-\exp (-\tau_{\rm HCN(4-3)})}{1-\exp (-\tau_{\rm HCO^+(4-3)})} 
\equiv \xi_{\rm HCN/HCO^+} \cdot \zeta_{\rm HCN/HCO^+},
\end{equation}
where $\xi_{\rm HCN/HCO^+}$ and $\zeta_{\rm HCN/HCO^+}$ correspond to 
the former and the latter term of the left hand side of Equation (\ref{eq6}), respectively. 
Also, we rewrite the optical depth as 
\begin{equation}\label{eq7}
\tau_{ul} = \frac{c^3}{8\pi \nu^3} \frac{g_u}{g_l} A_{ul} \frac{N_l}{dV} \left(1-\exp \left(\displaystyle - \frac{\Delta E_{ul}}{T_{\rm ex}}\right) \right),
\end{equation}
where $N_l$ is the line-of-sight column density at the lower energy level. 
The resultant $\xi_{\rm HCN/HCO^+}$ and $\zeta_{\rm HCN/HCO^+}$ are plotted in Figure \ref{figure7}(a) 
for the representative case of $N_{\rm HCO^+}/dV$ = 5 $\times$ 10$^{13}$ cm$^{-2}$ (km s$^{-1}$)$^{-1}$. 
At this $N_{\rm HCO^+}/dV$ with $X_{\rm HCN}/X_{\rm HCO^+}$ = 1, 
HCN(4-3) as well as HCO$^+$(4-3) are moderately optically thick, 
whereas HCN(4-3) can be {\it{heavily}} optically thick when $X_{\rm HCN}/X_{\rm HCO^+}$ = 10 (see also Figure \ref{figure4}). 
From Figure \ref{figure7}(a), one may find that $\xi_{\rm HCN/HCO^+}$ and $\zeta_{\rm HCN/HCO^+}$ 
vary towards the opposite direction at $T_{\rm bg}$ $\la$ 10 K 
(the range where both collisional and radiative processes can influence the excitation), 
which compensates with each other to keep the $R_{\rm HCN/HCO^+}$ more or less constant. 
At $T_{\rm bg}$ $\ga$ 10 K where $T_{\rm ex}$ $\sim$ $T_{\rm bg}$, on the other hand, 
both $\xi_{\rm HCN/HCO^+}$ and $\zeta_{\rm HCN/HCO^+}$ themselves are not so sensitive to $T_{\rm bg}$ anymore. 
Note that we can expect $R_{\rm HCN/HCO^+}$ $\sim$ 1 when both lines are thermalized and at the optically thick limit. 

As for the dependence on $T_{\rm kin}$, 
one can see that the $R_{\rm HCN/HCO^+}$ increases as $T_{\rm kin}$ gets higher in each panel, 
reflecting the higher $n_{\rm cr}$ of HCN(4-3) than HCO$^+$(4-3), 
except for the case of $X_{\rm HCN}$ = $X_{\rm HCO^+}$ in Figure \ref{figure6}(d). 
In that exceptional case, where both lines are quite optically thick, 
HCO$^+$(4-3) is fully thermalized because of the well reduced 
$n_{\rm cr,eff}$ ($\ll$ $n_{\rm H_2}$ = 10$^5$ cm$^{-3}$) 
due to the photon trapping (see also Appendix-\ref{app-B}), 
whereas HCN(4-3) is still sub-thermally excited, 
and $T_{\rm ex}/T_{\rm kin}$ of HCN(4-3) is maximized at $T_{\rm kin}$ = 50 K. 

Regarding the molecular abundance ratio, we find $X_{\rm HCN}/X_{\rm HCO^+}$ $\ga$ 10 is necessary 
to reproduce $R_{\rm HCN/HCO^+}$ $>$ 1 observed in AGNs 
in any $T_{\rm kin}$ and $N_{\rm mol}/dV$ studied here. 
Enhancing $X_{\rm HCN}$ (equivalently enhancing $N_{\rm HCN}$ in our models) will increase $\tau_{\rm HCN(4-3)}$ 
and $T_{\rm ex}$ (photon trapping), both of which will result in subsequent enhancement of the $R_{\rm HCN/HCO^+}$. 
The required $X_{\rm HCN}/X_{\rm HCO^+}$ is consistent with our previous multi-transitional 
non-LTE modeling of HCN and HCO$^+$ in NGC 1097 (\citetalias{2013PASJ...65..100I}). 
We should also mention that \citet{2007ApJ...671...73Y} concluded, 
based on their three-dimensional radiative transfer simulations, 
that $X_{\rm HCN}$ must be an order of magnitude higher than $X_{\rm HCO^+}$ 
in order to account for the observed high HCN(1-0)/HCO$^+$(1-0) ratios in AGNs 
(e.g., $\sim$ 2 in NGC 1068; \citealt{2008Ap&SS.313..279K}). 
Our results seem to be consistent with their modelings, although we here use $J$ = 4--3 transitions. 
Moreover and importantly, the $X_{\rm HCN}/X_{\rm HCO^+}$ required for AGNs ($\ga$ 10) 
are significantly higher than that required to reproduce the $R_{\rm HCN/HCO^+}$ 
in SB galaxies, which is typically $X_{\rm HCN}/X_{\rm HCO^+}$ $\sim$ 3. 
Therefore, the {\it{boosting factor}} of the abundance ratio 
in AGNs over that in SB galaxies is at least $\sim$ 3. 
This factor can even increase to $\ga$ 10 (i.e., $X_{\rm HCN}/X_{\rm HCO^+}$ $\ga$ 30) 
to account for the high-end values observed in AGNs 
(e.g., NGC 1068 (W-knot) in Table \ref{tbl1}) based on our modelings. 
For a convenient discussion, we here define the boosting factor as 
\begin{equation}\label{eq8}
BF_{\rm HCN/HCO^+} = \frac{(X_{\rm HCN}/X_{\rm HCO^+})_{\rm AGN}}{(X_{\rm HCN}/X_{\rm HCO^+})_{\rm SB}}, 
\end{equation}
where ($X_{\rm HCN}$/$X_{\rm HCO^+}$)$_{\rm AGN}$ and ($X_{\rm HCN}$/$X_{\rm HCO^+}$)$_{\rm SB}$ 
denote the molecular fractional abundance ratios in AGN and SB galaxies, respectively. 
The same notation is also used for the case of $X_{\rm HCN}$/$X_{\rm CS}$. 
Note that, for the extreme case like NGC 1068 (W-knot), 
we require $N_{\rm HCO^+}$/$dV$ $\la$ 5 $\times$ 10$^{13}$ cm$^{-2}$ (km s$^{-1}$)$^{-1}$ 
because $R_{\rm HCN/HCO^+}$ will eventually converge to unity 
for larger optical depths and never reaches such high values observed as $\sim$ 3. 

(ii) $R_{\rm HCN/CS}$---As is not the case of the $R_{\rm HCN/HCO^+}$, 
$R_{\rm HCN/CS}$ shows a steep dive in Figure \ref{figure6}(e)(f) as increasing $T_{\rm bg}$ from 2.73 K to 10 K, 
when $X_{\rm HCN}/X_{\rm CS}$ = 10. 
This feature is more prominent at lower $T_{\rm kin}$. 
All other cases show almost constant $R_{\rm HCN/CS}$ against $T_{\rm bg}$. 
Following the same manner in Equation (\ref{eq6}), we express the $R_{\rm HCN/CS}$ as 
\begin{equation}\label{eq9}
\frac{T_{\rm ex,HCN(4-3)} - T_{\rm bg}}{T_{\rm ex,CS(7-6)} - T_{\rm bg}} \cdot \frac{1-\exp (-\tau_{\rm HCN(4-3)})}{1-\exp (-\tau_{\rm CS(7-6)})} 
\equiv \xi_{\rm HCN/CS} \cdot \zeta_{\rm HCN/CS}.
\end{equation}
The resultant $\xi_{\rm HCN/CS}$ and $\zeta_{\rm HCN/CS}$ are plotted in Figure \ref{figure7}(b) 
for the case of $N_{\rm CS}/dV$ = 5 $\times$ 10$^{13}$ cm$^{-2}$ (km s$^{-1}$)$^{-1}$ as a representative example. 

In Figure \ref{figure7}(b), a rapid drop in $\zeta_{\rm HCN/CS}$ 
(by a factor of $\sim$ 5 when $X_{\rm HCN}/X_{\rm CS}$ = 10) stands out, 
which damps the variation of $\xi_{\rm HCN/CS}$. 
We can attribute this dive to the quite different $\tau$ between HCN(4-3) and CS(7-6); 
$\tau_{\rm HCN(4-3)}$ = 19.5-41.8, whereas $\tau_{\rm CS(7-6)}$ = 0.06-0.45 
for the case of Figure \ref{figure7}(b). 
In this case, $\zeta_{\rm HCN/CS}$ mostly reflects the variation of $\tau_{\rm CS(7-6)}$ 
since $\zeta_{\rm HCN/CS}$ is now $\sim$ $1/\tau_{\rm CS(7-6)}$. 
Then, considering the $\tau$ of HCN(4-3) and CS(7-6) shown in Figure \ref{figure4}, 
we can deduce that the condition of, {\it{$``$HCN(4-3) is optically thick, whereas CS(7-6) is optically thin,$"$}} 
would be the key to realize the high $R_{\rm HCN/CS}$ $\ga$ 10 observed in some AGNs. 
This scenario can explain the dependence of $R_{\rm HCN/CS}$ on $T_{\rm kin}$, 
as we can expect larger $\tau_{\rm CS(7-6)}$ (higher $T_{\rm ex,CS(7-6)}$) at higher $T_{\rm kin}$, 
while $\tau_{\rm HCN(4-3)}$ is already substantially large 
regardless of any $T_{\rm kin}$ in our models. 
Note that $R_{\rm HCN/CS}$ naturally converges to $\sim$ 1 
as both HCN(4-3) and CS(7-6) reach optically thick limit and fully thermalized. 

Then, we suggest that $N_{\rm CS}/dV$ $\la$ 5 $\times$ 10$^{13}$ cm$^{-2}$ (km s$^{-1}$)$^{-1}$ 
and $X_{\rm HCN}/X_{\rm CS}$ $\ga$ 10 would be necessary conditions 
to reproduce the observed high $R_{\rm HCN/CS}$ in AGNs. 
These conditions become more tighter ones 
if AGNs have $T_{\rm bg}$ $\ga$ 20 K and $T_{\rm kin}$ $\ga$ 100 K (Figure \ref{figure6}). 
The former assumption can be justified by the high dust temperature 
in NGC 1068 (46 K; \citealt{2014A&A...567A.125G}). 
As for the latter one, it has been suggested that $T_{\rm kin}$ in AGN environments is 
as high as several hundreds K even in a molecular phase 
(e.g., \citetalias{2013PASJ...65..100I}; \citealt{1998ApJ...495..267M,2008ApJ...677..262K,2012A&A...537A.133D,2014A&A...570A..28V}), 
thus is highly likely to be satisfied. 
The required $X_{\rm HCN}/X_{\rm CS}$ for AGNs ($\ga$ 10) is 
again significantly higher than that for SB galaxies ($\sim$ 3). 
Hence the estimated $BF_{\rm HCN/CS}$ is at least $\sim$ 3. 
This $BF_{\rm HCN/CS}$ can increase to $\sim$ 10 (i.e., $X_{\rm HCN}$/$X_{\rm CS}$ $\sim$ 30 in AGNs) 
to reproduce the high-end values of $R_{\rm HCN/CS}$ $\ga$ 12 (Figure \ref{figure1}) 
under the conditions of moderately high dust temperature of $\ga$ 10 K and high $T_{\rm kin}$ such as 200 K. 
These boosting factors are roughly consistent with $BF_{\rm HCN/HCO^+}$. 

\subsection{Dependence of the line ratios on gas density}\label{sec4.4}
\begin{figure*}
\epsscale{1.2}
\plotone{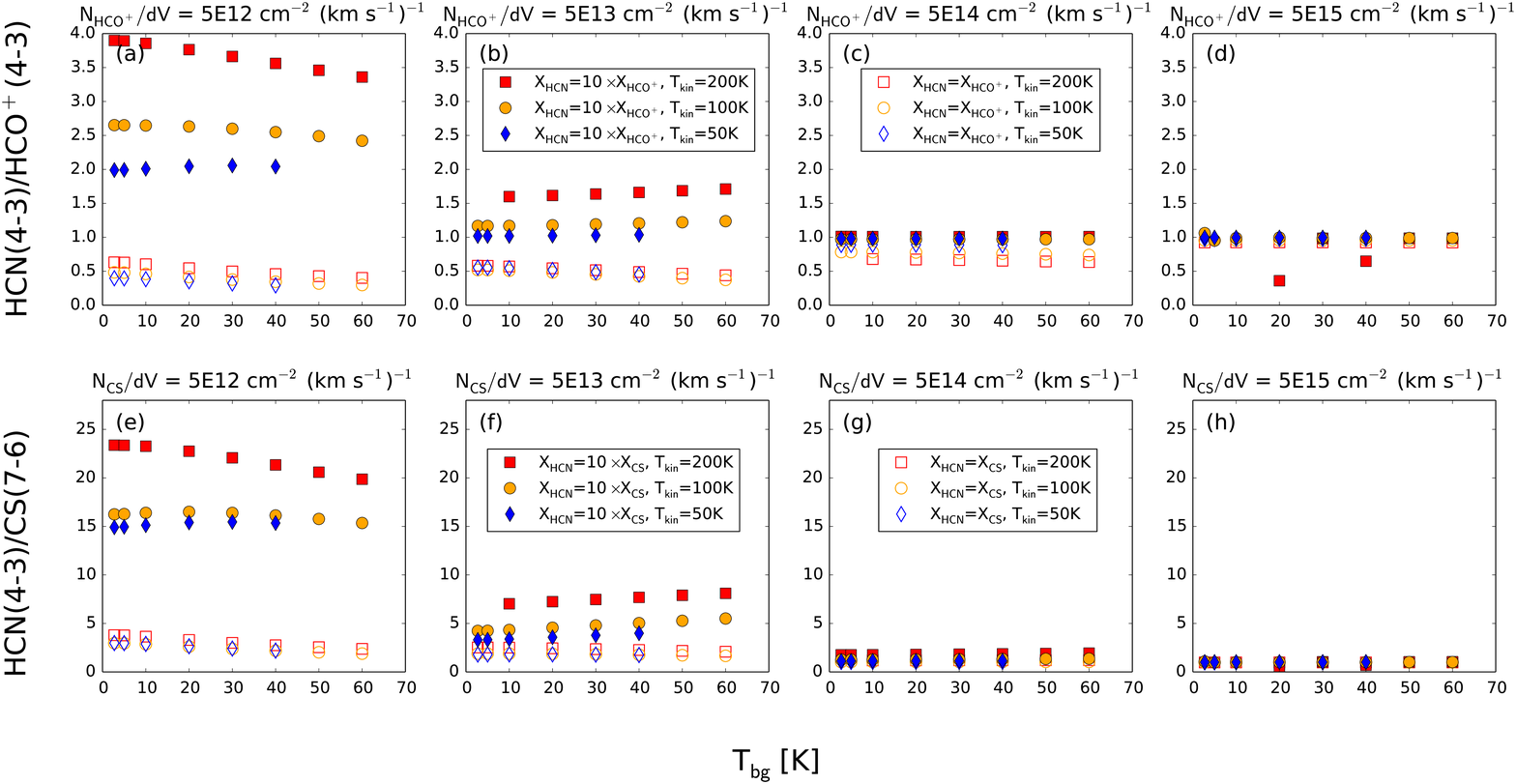}
\caption{
Same as Figure \ref{figure6}, but the models with $n_{\rm H_2}$ = 5 $\times$ 10$^6$ cm$^{-3}$ are shown. 
}
\label{figure8}
\end{figure*}

\begin{figure*}
\epsscale{1.2}
\plotone{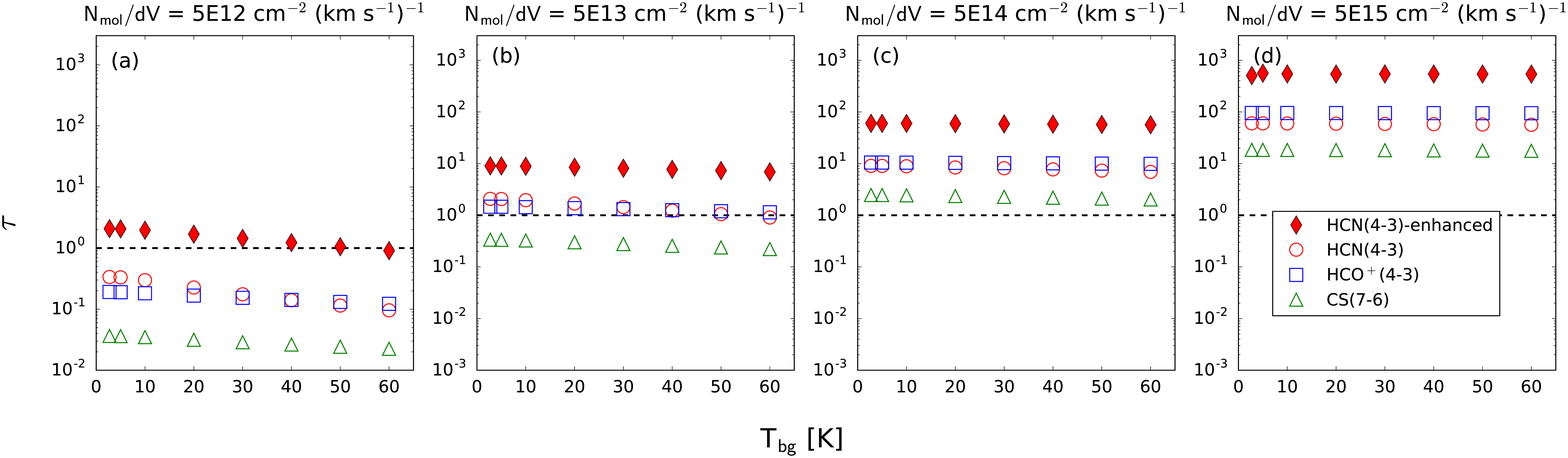}
\caption{
Same as Figure \ref{figure4}, but the models with $n_{\rm H_2}$ = 5 $\times$ 10$^6$ cm$^{-3}$ are shown. 
The dashed line indicates $\tau$ = 1.0 for an eye-guide. 
}
\label{figure9}
\end{figure*}

Recently, \citet{2014A&A...570A..28V} reported that $n_{\rm H_2}$ 
would vary even inside a single CND around an AGN. 
Therefore, the impact of different densities on the line ratios of our interest 
should be investigated as a subsequent analysis to Section \ref{sec4.2}-\ref{sec4.3}. 
To this end, we conducted the same analysis as shown in the previous parts 
but for the case of $n_{\rm H_2}$ = 5 $\times$ 10$^6$ cm$^{-3}$. 
This density roughly corresponds to the highest one predicted 
by \citet{2014A&A...570A..28V} in the CND of NGC 1068. 
The resultant (model-predicted) line ratios and optical depths 
are presented in Figures \ref{figure8} and \ref{figure9}, respectively. 

From that Figure, we found that both $R_{\rm HCN/HCO^+}$ and $R_{\rm HCN/CS}$ tend to show 
higher values than those in Figure \ref{figure6} when 
$N_{\rm mol}$/$dV$ $\leq$ 5 $\times$ 10$^{13}$ cm$^{-2}$ (km s$^{-1}$)$^{-1}$ (mol = HCO$^+$ and CS), 
i.e., cases when emission lines are optically thin or moderately thick. 
Especially, $R_{\rm HCN/HCO^+}$ in Figure \ref{figure8}(a) 
exceeds 2.5 at $T_{\rm kin}$ $\geq$ 100 K and $X_{\rm HCN}$/$X_{\rm HCO^+}$ = 10, 
which is comparable to the observed high-end ratios in Table \ref{tbl1}. 
The higher values than the former cases with $n_{\rm H_2}$ = 10$^5$ cm$^{-3}$ 
directly reflects the highest $n_{\rm cr}$ of HCN(4-3) among the lines used here. 
However, when $N_{\rm mol}$/$dV$ is increased, 
one can also see that the line ratios converge to unity more quickly 
than for the cases in Figure \ref{figure6}. 
Enhanced excitation due to the higher $n_{\rm H_2}$ makes each emission line 
being thermalized and optically thick, leading to this convergence. 
This is also manifested in the trend that both line ratios are less dependent on $T_{\rm bg}$, 
but more sensitive to $T_{\rm kin}$ than the cases in Figure \ref{figure6}, 
which is especially prominent in $R_{\rm HCN/CS}$ (this ratio is sensitive to the optical depth; Section \ref{sec4.3}). 
Therefore, although it depends on the line opacity, 
higher $n_{\rm H_2}$ does not necessarily correspond to higher $R_{\rm HCN/HCO^+}$ nor $R_{\rm HCN/CS}$. 

Then, let us speculate more in detail about three cases of gas densities in the following: 
($i$) $n_{\rm AGN}$ $=$ $n_{\rm SB}$, 
($ii$) $n_{\rm AGN}$ $<$ $n_{\rm SB}$, and 
($iii$) $n_{\rm AGN}$ $>$ $n_{\rm SB}$. 
Here, $n_{\rm AGN}$ and $n_{\rm SB}$ denote {\it{representative}} gas densities in 
AGN and SB environments, respectively. 
We focus on two cases of $n_{\rm H_2}$ 
as are shown in Figures \ref{figure6} (10$^5$ cm$^{-3}$) 
and \ref{figure8} (5 $\times$ 10$^6$ cm$^{-3}$), 
and also use only $R_{\rm HCN/HCO^+}$ during this speculation, for simplicity. 
Then, the higher density between $n_{\rm AGN}$ and $n_{\rm SB}$ 
corresponds to 5 $\times$ 10$^6$ cm$^{-3}$, 
whereas the lower one is 10$^5$ cm$^{-3}$ hereafter. 
Following the same manner, an essentially similar argument for abundance ratios 
can be achieved for the case of $X_{\rm HCN}$/$X_{\rm CS}$ as well. 
Note that, however, we can only argue for the rough trend 
of relative difference in the abundance ratio 
between AGNs and SB galaxies in the following, 
because of this simple treatment. 
More comprehensive and quantitative comparison requires 
high resolution, multi-line, and multi-species analysis to restrict the gas excitation. 
Such an analysis should be conducted with future observations. 

($i$) $n_{\rm AGN}$ $=$ $n_{\rm SB}$: 
in the case of $n_{\rm H_2}$ = 10$^5$ cm$^{-3}$ (Figure \ref{figure6}), 
$X_{\rm HCN}$/$X_{\rm HCO^+}$ is $\sim$ a few for SB galaxies (Section \ref{sec4.3}). 
Then, we request for the boosting factor $BF_{\rm HCN/HCO^+}$ to be at least $\sim$ 3
to explain the observed $R_{\rm HCN/HCO^+}$ in AGNs. 
Regarding the high-end value such as $R_{\rm HCN/HCO^+}$ $\ga$ 3 (e.g., NGC 1068 (W-knot)), 
$BF_{\rm HCN/HCO^+}$ even reaches $\ga$ 10 (i.e., $X_{\rm HCN}$/$X_{\rm HCO^+}$ $\ga$ 30). 
When $n_{\rm H_2}$ is increased to 5 $\times$ 10$^6$ cm$^{-3}$ (Figure \ref{figure8}), 
$X_{\rm HCN}$/$X_{\rm HCO^+}$ = 1 
is sufficient to reproduce the $R_{\rm HCN/HCO^+}$ in SB galaxies (Figure \ref{figure8}). 
On the other hand, we still need $BF_{\rm HCN/HCO^+}$ to be a few 
to even $\sim$ 10 (depending on the line opacity) 
in order to account for the $R_{\rm HCN/HCO^+}$ in AGNs, especially the high-end values. 
In either case, we require the enhanced HCN abundance 
in AGNs to reproduce the observations. 

($ii$) $n_{\rm AGN}$ $<$ $n_{\rm SB}$: 
As shown above, $X_{\rm HCN}$/$X_{\rm HCO^+}$ $\sim$ 1 
is sufficient to reproduce the observed $R_{\rm HCN/HCO^+}$ in SB galaxies 
when $n_{\rm H_2}$ = 5 $\times$ 10$^6$ cm$^{-3}$ (Figure \ref{figure8}). 
However, we need to enhance the $X_{\rm HCN}$/$X_{\rm HCO^+}$ up to $\sim$ 10 
($\ga$ 30 for the case of the high-end value) to explain the $R_{\rm HCN/HCO^+}$ in AGNs 
when $n_{\rm H_2}$ = 10$^5$ cm$^{-3}$. 
Hence, again it indicates substantially high $X_{\rm HCN}$/$X_{\rm HCO^+}$ in AGNs with 
$BF_{\rm HCN/HCO^+}$ of as high as $\ga$ 10--30. 

($iii$) $n_{\rm AGN}$ $>$ $n_{\rm SB}$: 
in the case of $n_{\rm H_2}$ = 10$^5$ cm$^{-3}$, 
$X_{\rm HCN}$/$X_{\rm HCO^+}$ $\sim$ 3 is sufficient 
to reproduce the observed $R_{\rm HCN/HCO^+}$ in SB galaxies (Figure \ref{figure6}). 
For AGNs having higher gas density of $n_{\rm H_2}$ = 5 $\times$ 10$^6$ cm$^{-3}$ (Figure \ref{figure8}), 
we have to restrict ourselves to $N_{\rm HCO^+}$/$dV$ $\la$ 5 $\times$ 10$^{13}$ cm$^{-2}$ (km s$^{-1}$)$^{-1}$ 
at first because $R_{\rm HCN/HCO^+}$ converges to unity for higher column density (or $\tau$) 
and thus never reaches $\ga$ 2. 
Under these conditions, we found $X_{\rm HCN}$/$X_{\rm HCO^+}$ $\sim$ 3--5 can indeed yield a high 
$R_{\rm HCN/HCO^+}$ of $\sim$ 1.5--2.5 especially when $N_{\rm HCO^+}$/$dV$ $\sim$ 
5 $\times$ 10$^{12}$ cm$^{-2}$ (km s$^{-1}$)$^{-1}$ and $T_{\rm kin}$ $\ga$ 100 K (Figure \ref{figure8}(a)). 
In this case, $BF_{\rm HCN/HCO^+}$ is only $\sim$ 1--1.5, 
i.e., no abundance variation between AGNs and SB galaxies. 
Note that, even with this high $n_{\rm H_2}$, it is still a bit challenging 
to reproduce the observed highest $R_{\rm HCN/HCO^+}$ of $\ga$ 3. 
To do so, we might need to require $BF_{\rm HCN/HCO^+}$ 
to be $\sim$ a few even with the high $n_{\rm H_2}$ in AGNs. 
On the other hand, when $N_{\rm HCO^+}$/$dV$ 
$\sim$ 5 $\times$ 10$^{13}$ cm$^{-2}$ (km s$^{-1}$)$^{-1}$, 
we can not avoid increasing $BF_{\rm HCN/HCO^+}$ $\ga$ a few 
to yield the high-end $R_{\rm HCN/HCO^+}$ in AGNs (Figure \ref{figure8}(b)). 

As the summary of our non-LTE modelings with both 
$n_{\rm H_2}$ = 10$^5$ cm$^{-3}$ and 5 $\times$ 10$^6$ cm$^{-3}$, 
we suggest that $X_{\rm HCN}/X_{\rm HCO^+}$ would be enhanced 
in AGNs as compared to SB galaxies by several times 
to $\ga$ 10 times when $n_{\rm H_2}$ in AGNs are comparable to or lower than those in SB galaxies 
in order to reproduce their observed $R_{\rm HCN/HCO^+}$. 
Another plausible origin might be the systematically higher gas density in AGNs than in SB galaxies, 
which should be coupled to the low opacity of the emission lines from AGNs, to yield the high $R_{\rm HCN/HCO^+}$. 
In this case, an almost comparable $X_{\rm HCN}/X_{\rm HCO^+}$ 
is expected between AGNs and SB galaxies. 
However, the feasibility of such a systematic difference in gas density is unclear at this moment. 
Although \citet{2014A&A...570A..28V} suggested such differences 
between the kpc scale SB ring and the 100 pc scale CND of NGC 1068, 
what we are treating here is the rather smaller, 
{\it{circumnuclear scale (likely to be $\la$ 100 pc scale)}} components 
located at the centers of both AGN and SB galaxies; 
recall that we made Figure \ref{figure1} only based on 
the high resolution sample in Table \ref{tbl1}. 
Hence, the contrast in density between AGN and circumnuclear SB environments 
would be smaller than in the case between AGN and kpc scale SB environments. 
Moreover, $\sim$ 30 pc scale measurements of HCN(4-3) and HCN(1-0) revealed 
HCN(4-3)/HCN(1-0) integrated intensity ratio is $\sim$ 0.7--1.0 in the central region of NGC 253 
(the brightness temperature scale, \citealt{2011ApJ...735...19S,2015ApJ...801...63M}), 
which value is comparable to those obtained 
within the CND of NGC 1068 (\citealt{2014A&A...570A..28V}). 
\citet{2007ApJ...666..156K} also proposed that 
the line excitation is independent of galaxy types, 
although this argument was based on {\it{single dish}} 
measurements of the HCN spectral energy distribution (SLED). 
These results suggest that the degree of line excitation 
at the nuclear region of a galaxy would not depend on the specific type of that galaxy. 
Likely higher $T_{\rm kin}$ in AGNs would lead to lower $n_{\rm H_2}$ than SB galaxies 
if they share the same line excitation. 
Therefore, we hereafter prefer the scenario that the variation of 
underlying molecular abundances between AGNs and SB galaxies 
as the prime cause of the HCN-enhancement. 
However, we should mention that we can not discard the possibility of 
systematically higher $n_{\rm H_2}$ in AGNs than in SB galaxies, 
nor the possibility that the CND of NGC 1068 has specifically and remarkably high $n_{\rm H_2}$ 
as compared to other AGNs and SB galaxies, 
because of the simplified analysis and discussion in this work. 

\subsection{Comparison with real galaxies: \\
Case study of NGC 1097, NGC 1068, and NGC 4418}\label{sec4.5} 
In this Section, we briefly compare our model calculations 
with the observed $R_{\rm HCN/HCO^+}$ and/or $R_{\rm HCN/CS}$ 
of two instructive AGNs and one buried-AGN, namely NGC 1097, NGC 1068, and NGC 4418. 
The characteristics of these galaxies are presented in Appendix-\ref{app-A}. 
Models with $n_{\rm H_2}$ = 10$^5$ cm$^{-3}$ (Figure \ref{figure6}) are employed here, 
but those with $n_{\rm H_2}$ = 5 $\times$ 10$^6$ cm$^{-3}$ yield essentially the same (qualitative) arguments. 
As these models were not constructed to mimic the environment of a specific galaxy, 
our immediate objective here is to roughly check the consistency of the trend 
between the model-predicted line ratios and the actual observed values. 

For NGC 1097 (AGN), \citet{2008ApJ...683...70H} estimated 
$N_{\rm H_2}$ = 5.9 $\times$ 10$^{22}$ cm$^{-2}$ from CO interferometric observations. 
Assuming $X_{\rm CS}$ = 2.5 $\times$ 10$^{-9}$ (the value of OMC-1 Extended Ridge where 
$N_{\rm H_2}$ = 3 $\times$ 10$^{23}$ cm$^{-2}$; \citealt{1987ApJ...315..621B}) 
and the line width of 250 km s$^{-1}$ (FWZI of HCN(4-3); \citetalias{2013PASJ...65..100I}), 
$N_{\rm CS}/dV$ is estimated to be $\sim$ 6 $\times$ 10$^{11}$ cm$^{-2}$ (km s$^{-1}$)$^{-1}$. 
Then, judging from Figure \ref{figure4}, we can expect CS(7-6) emission 
is totally optically thin ($\tau_{\rm CS(7-6)}$ $\ll$ 0.1). 
This is consistent with the non-detection of CS(7-6) emission 
in our previous observations with ALMA (\citetalias{2013PASJ...65..100I}). 
Even if we increase $X_{\rm CS}$ by 10 times, the situation is the same. 
On the other hand, non-LTE analysis by \citetalias{2013PASJ...65..100I} suggested 
$N_{\rm HCN}/dV$ in NGC 1097 is $\sim$ a few $\times$ 10$^{13}$ cm$^{-2}$ (km s$^{-1}$)$^{-1}$, 
thus HCN(4-3) would be moderately optically thick (Figure \ref{figure4}). 
The resultant $X_{\rm HCN}$ in this case is $\sim$ 10$^{-7}$, 
which is close to the value observed in Galactic hot cores 
and is significantly higher than that observed in, e.g., OMC-1 Extended Ridge (5.0 $\times$ 10$^{-9}$; \citealt{1987ApJ...315..621B}). 
Therefore, the situation is similar to the panel-(e) of Figure \ref{figure6} 
with a high $X_{\rm HCN}/X_{\rm CS}$ of $\ga$ 10. 
Indeed, our model using the above values shows $R_{\rm HCN/CS}$ $\ga$ 20, 
which matches the observed value ($R_{\rm HCN/CS}$ $>$ 12.7). 
Furthermore, an optical depth of HCN(1-0) estimated from the above number is 
$\tau_{\rm HCN(1-0)}$ $\ga$ 0.1 ($T_{\rm kin}$ = 100 K is assumed), 
which is consistent with the estimation by \citet{2015A&A...573A.116M}. 

However, our estimated abundance ratios such as $X_{\rm HCN}$/$X_{\rm HCO^+}$ $\sim$ 10 
are significantly higher than those obtained by \citet{2015A&A...573A.116M} through LTE analysis, which are $\sim$ 3. 
As long as we adopt a gas density that is not high enough to thermalize 
all of HCN(4-3), HCO$^+$(4-3), and CS(7-6) emission lines (e.g.., $n_{\rm H_2}$ = 10$^5$ cm$^{-3}$ in Figure \ref{figure6}), 
we should require highly enhanced abundance ratios to overcome the inefficient excitation 
of the HCN(4-3) line (this line has the highest $n_{\rm cr}$ among the target lines; Table \ref{tbl2}) 
and yield the high line ratios observed. 
In the case of higher $n_{\rm H_2}$ such as 5 $\times$ 10$^6$ cm$^{-3}$, 
the required abundance ratio can be as low as $\sim$ 3 for optically thin emission lines (Figure \ref{figure8}-(a)(e)). 
This ratio is consistent with the measured values by \citet{2015A&A...573A.116M}, 
because now the conditions of ($i$) LTE and ($ii$) optically thin emission are mostly satisfied. 
For optically thicker cases, we will again see discrepancy 
between our model predictions and those by \citet{2015A&A...573A.116M}. 
Therefore, whether our model-predictions are consistent or not with the previous LTE results 
strongly depends on the assumed $n_{\rm H_2}$ and line optical depth. 

In the case of NGC 1068, both the ratios of $R_{\rm HCN/HCO^+}$ and $R_{\rm HCN/CS}$ 
at the precise AGN position are lower than those of NGC 1097 (AGN). 
We suppose this is due likely to a larger $N_{\rm mol}$ than that of NGC 1097 (AGN), 
i.e., the lines are optically thicker than those of NGC 1097 (AGN). 
Actually, the $N_{\rm H_2}$ towards this AGN can be $\sim$ 6 $\times$ 10$^{23}$ cm$^{-2}$ 
by using the results of the CO multi-line non-LTE analysis by \citet{2014A&A...570A..28V} 
(see also the footnote in Section \ref{sec2} for the estimation), 
which is $\sim$ 10 times larger than that of NGC 1097 (AGN). 
In this case, again by applying the $X_{\rm CS}$ = 2.5 $\times$ 10$^{-9}$ (\citealt{1987ApJ...315..621B}) 
and $dV$ $\sim$ 200 km s$^{-1}$ (\citealt{2014A&A...567A.125G}), 
we estimate $N_{\rm CS}$/$dV$ $\sim$ 1 $\times$ 10$^{13}$ cm$^{-2}$ (km s$^{-1}$)$^{-1}$. 
For this $N_{\rm CS}$/$dV$, CS(7-6) emission is still moderately optically thin ($\tau_{\rm CS}$ $\sim$ 0.1; Figure \ref{figure4}), 
but is significantly optically thicker than the case of NGC 1097 (AGN). 
The same speculation can also be applied to HCN(4-3) and HCO$^+$(4-3) emission lines. 
As shown in Figure \ref{figure6}, it is natural for line ratios to converge to unity when the paired lines get optically thicker. 
Moreover, since $R_{\rm HCN/CS}$ is sensitive to $T_{\rm bg}$, we can expect lower value of this ratio 
in NGC 1068 (AGN) than in NGC 1097 (AGN) even when they share the same $X_{\rm HCN}$/$X_{\rm CS}$ 
considering the much higher AGN luminosity of the former than the latter (\citealt{2012ApJ...748..130M,2014ApJ...783..106L}). 
Therefore, despite the rather lower $R_{\rm HCN/HCO^+}$ and $R_{\rm HCN/CS}$ than NGC 1097 (AGN), 
we still expect underlying $X_{\rm HCN}/X_{\rm HCO^+}$ 
and $X_{\rm HCN}/X_{\rm CS}$ are high ($\sim$ 10) in NGC 1068 (AGN). 
Note that the line optical depths of HCN(4-3) are 40--60 even when 
$N_{\rm HCN}$/$dV$ = 1 $\times$ 10$^{15}$ cm$^{-2}$ (km s$^{-1}$)$^{-1}$, 
which are still smaller than that observed in Arp 220W (a system of self-absorption; \citealt{2015ApJ...800...70S}). 

As for the {\it{knots}} of the CND of NGC 1068, 
panels-(a)(b)(e)(f) of Figure \ref{figure6} would be good approximations for their situations 
if we adopt the estimated column densities by \citet{2014A&A...570A..28V}. 
In those cases with $T_{\rm kin}$ $>$ 100 K, we need 
the abundance ratios to be $\ga$ 30 
when $n_{\rm H_2}$ = 10$^5$ cm$^{-3}$. 
As stated before, we can not discard the possibility of no abundance enhancement 
in the case of the panels-(a) and (e) in Figure \ref{figure8}. 
However, if $N_{\rm mol}$/$dV$ $\ga$ 5 $\times$ 10$^{13}$ cm$^{-2}$ (km s$^{-1}$)$^{-1}$ is true, 
we still need to enhance the abundance ratios to be $\ga$ 30 
even with $n_{\rm H_2}$ = 5 $\times$ 10$^6$ cm$^{-3}$ 
to reproduce the high-end values observed in these knots (e.g., $R_{\rm HCN/HCO^+}$ $\ga$ 3). 
Note that we can find consistent abundance ratios with those suggested above 
in previous works of multi-line, multi-species non-LTE analysis (e.g., Table 8 of \citealt{2014A&A...570A..28V}). 
On the other hand, there seems to be an inconsistency with the predicted abundance ratios 
from LTE analysis (Table 5 of \citealt{2014A&A...570A..28V}), 
which are $\sim$ 5 for both $X_{\rm HCN}$/$X_{\rm HCO^+}$ and $X_{\rm HCN}$/$X_{\rm CS}$. 
For $n_{\rm H_2}$ of 10$^5$ cm$^{-3}$ (Figure \ref{figure6}), 
again we should enhance $X_{\rm HCN}$ to overcome the inefficient excitation of HCN(4-3). 
For the higher $n_{\rm H_2}$ of 5 $\times$ 10$^6$ cm$^{-3}$ (Figure \ref{figure8}), 
our results are consistent with those from the LTE analysis 
as long as emission lines are optically thin or moderately thick (see also Figure \ref{figure9}), 
but the discrepancy becomes prominent for optically thicker cases 
because now the condition of {\it{optically thin emission}} assumed in the LTE analysis (\citealt{2014A&A...570A..28V}) breaks. 
In such cases, we eventually require enhanced, e.g., $X_{\rm HCN}$/$X_{\rm HCO^+}$ of $\ga$ 30 as already addressed. 

In the case of NGC 4418, the same argument of the optical depth effect 
as for NGC 1068 (AGN) holds with a stronger basis, 
because this galaxy hosts a heavily obscured compact nucleus 
($N_{\rm H_2}$ $\ga$ 10$^{26}$ cm$^{-2}$; \citealt{2013ApJ...764...42S}). 
The distinctively lower $R_{\rm HCN/CS}$ than {\it{normal}} AGNs 
favors this view as this ratio is highly sensitive to 
the optical depth of CS(7-6) emission line (Figure \ref{figure7}). 
Regarding $R_{\rm HCN/HCO^+}$, we mention this ratio is slightly increased to $\sim$ 2 when observed 
at 0$''$.5 $\sim$ 80 pc resolution (\citealt{2013ApJ...764...42S}), 
which is a comparable value to other AGNs. 
However, whether an enhanced $X_{\rm HCN}$/$X_{\rm HCO^+}$ such as 10 
is really necessary to account for this relatively high ratio is unclear, 
because this galaxy exhibits prominent $v$ = 1 state 
HCN emission line (IR-pumping; \citealt{2010ApJ...725L.228S}). 
Radiative transfer modelings involving the IR-pumping are thus inevitable to 
elucidate the validity of our diagram in such a deeply embedded nucleus 
that the IR-pumping must affect the ratio.

\subsection{Implication to generic HCN-diagrams}\label{sec4.6}
\begin{figure*}
\epsscale{1}
\plotone{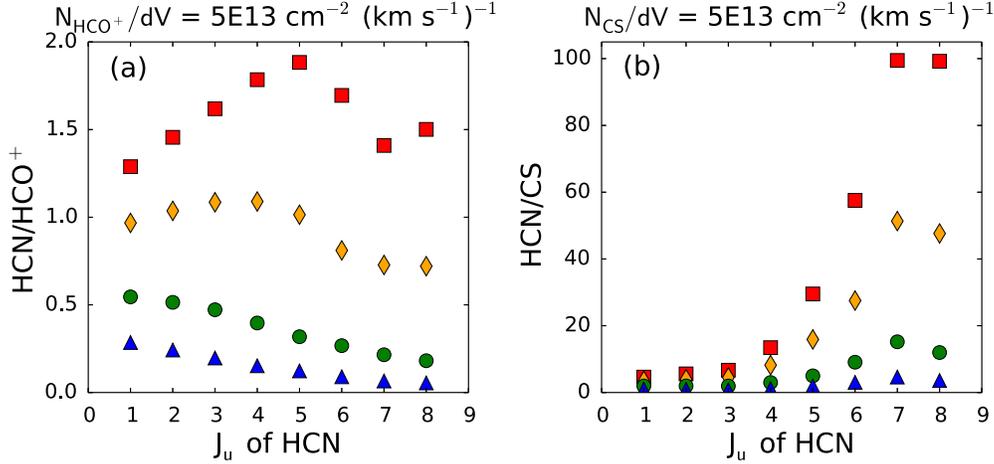}
\caption{
An example of spectral ratio distribution of (a) HCN to HCO$^+$ and (b) HCN to CS integrated intensity ratios 
in the brightness temperature scale, as a function of the upper rotational state $J_{\rm u}$ of HCN. 
The value of each ratio is calculated based on our RADEX simulation, assuming $n_{\rm H_2}$ = 10$^5$ cm$^{-3}$, 
$T_{\rm kin}$ = 100 K, $T_{\rm bg}$ = 10 K. 
The case of $N_{\rm HCO^+}/dV$ (or $N_{\rm CS}/dV$) = 5 $\times$ 10$^{13}$ cm$^{-2}$ (km s$^{-1}$)$^{-1}$ is shown. 
We set molecular abundance ratios to be $X_{\rm HCN}/X_{\rm HCO^+}$ (or $X_{\rm HCN}/X_{\rm CS}$) 
= 20 (red square), 10 (orange diamond), 3 (green circle), and 1 (blue triangle), respectively. 
As for CS, the transition with the closest frequency to each HCN($J_u$-$J_l$) is used. 
At any transition, enhanced $X_{\rm HCN}$ correspondingly produces higher line ratios (i.e., blue $<$ green $<$ orange $<$ red). 
}
\label{figure10}
\end{figure*}

As the last of Section \ref{sec4}, we here predict HCN-to-HCO$^+$ and HCN-to-CS integrated intensity ratios 
at other transitions based on our RADEX modelings, 
by varying $X_{\rm HCN}/X_{\rm HCO^+}$ and $X_{\rm HCN}/X_{\rm CS}$. 
Illustrative cases are shown in Figure \ref{figure10} 
as a function of the upper rotational state ($J_{\rm u}$) of HCN. 
As for CS, we coupled a transition with the closest frequency to each HCN transition (e.g., HCN(1-0) and CS(2-1), HCN(4-3) and CS(7-6)). 
We hereafter call this type of diagram as a spectral ratio distribution ({\it{SRD}}). 
Note that we do not intend to predict a line ratio of a specific galaxy in Figure \ref{figure10}, 
but to understand a qualitative feature of these ratios at various transitions. 

By inspecting the SRD, one can see that enhanced 
$X_{\rm HCN}/X_{\rm HCO^+}$ (or $X_{\rm HCN}/X_{\rm CS}$) 
correspondingly produces higher line ratios at any transition. 
Therefore, if $X_{\rm HCN}$/$X_{\rm HCO^+}$ and $X_{\rm HCN}$/$X_{\rm CS}$ 
are truly enhanced in AGNs than in SB galaxies, and if they share the comparable excitation (e.g., $n_{\rm H_2}$ and $T_{\rm kin}$), 
we predict AGNs would show higher line ratios not only at a single transition 
but also at other transitions than SB galaxies. 
Indeed, NGC 1097 and NGC 1068 show higher HCN(1-0)/HCO$^+$(1-0) and HCN(1-0)/CS(2-1) ratios 
than SB galaxies (e.g., \citealt{2008Ap&SS.313..279K,2011ApJ...728L..38N,2015A&A...573A.116M}). 

From this perspective, remarkable inconsistency can stem from NGC 7469, 
which shows HCN(1-0)/HCO$^+$(1-0) ratio of $\sim$ 0.6 
when observed with Nobeyama Millimeter Array ($\sim$ 6$''$ = 2 kpc aperture; \citealt{2005AIPC..783..203K}), 
whereas it shows HCN(4-3)/HCO$^+$(4-3) = 1.1 ($\sim$ 150 pc aperture; Table \ref{tbl1}). 
This inconsistency can be reconciled when we measure $J$ = 1--0 lines at a higher resolution 
to selectively probe the CND of NGC 7469 as much as possible, 
because this galaxy hosts a prominent SB ring with $\sim$ 1 kpc in diameter 
that can be a strong source of spectral contamination; 
HCN(4-3)/HCO$^+$(4-3) ratio at the SB ring is as low as $\sim$ 0.5. 
In fact, $\sim$ 150 pc aperture measurements of $J$ = 1--0 transitions 
newly obtained with ALMA reveals HCN(1-0)/HCO$^+$(1-0) $>$ 1 at the AGN position (Izumi et al. in preparation). 
This improved spatial resolution is still larger than, but more close to 
the expected size of the XDR in NGC 7469 ($\sim$ 80 pc in diameter; \citealt{2015ApJ...811...39I}). 
With this in mind, we here suppose that the low HCN(1-0)/HCO$^+$(1-0) ratio ($<$ 1) 
observed in NGC 2273 and NGC 4051 (\citealt{2012MNRAS.424.1963S}) 
would be (at least partly) due to severe spectral contamination from the surrounding SB regions. 
This can be the same situation as we discussed for NGC 1365 and NGC 4945 in Section \ref{sec3}. 
These low line ratios were measured in the central 3$''$ regions (\citealt{2012MNRAS.424.1963S}), 
which correspond to 390 pc for NGC 2273 and 150 pc for NGC 4051, respectively. 
Indeed, the equivalent widths of the 11.3 $\mu$m PAH emission 
in NGC 2273 (330 nm with $\sim$ 500 pc slit; \citealt{2014MNRAS.443.2766A}) 
and NGC 4051 (95 nm with 25 pc slit; \citealt{2016MNRAS.455..563A}) 
are significantly larger than those of NGC 7469 or NGC 1068 (Section \ref{sec3}). 
Moreover, the 2-10 keV X-ray luminosity of NGC 2273 
(log ($L_{\rm 2-10keV}$/erg s$^{-1}$) = 42.7; \citealt{2012ApJ...748..130M}) 
and NGC 4051 (log ($L_{\rm 2-10keV}$/erg s$^{-1}$) = 41.1; \citealt{2014ApJ...783..106L}) 
are $\sim$ 3 and $\sim$ 130 times smaller than that of NGC 7469 
(log ($L_{\rm 2-10keV}$/erg s$^{-1}$) = 43.2; \citealt{2014ApJ...783..106L}), respectively. 
Therefore, we can expect that the extent of the hypothesized XDRs 
are much smaller in NGC 2273 and NGC 4051 than that of NGC 7469 (\citealt{2015ApJ...811...39I}), 
hence, than the above mentioned 3$''$ areas where the line ratios were measured. 
On the other hand, again we cannot discard the possibility 
that these galaxies have less dense gas 
than AGNs with high line ratios (see also Section \ref{sec4.4}). 
Multi-line, multi-species modelings are indeed required to disentangle these scenarios. 

\section{Possible chemical causes for the molecular abundance variation}\label{sec5}
In this last Section, we present various scenarios for 
the molecular abundance variation 
(or, HCN-enhancement in AGNs) as suggested in Section \ref{sec4}, 
from the perspective of ISM chemistry. 
We here focus on the $R_{\rm HCN/HCO^+}$ 
of AGNs and SB galaxies for simplicity, 
since this ratio is less sensitive to excitation than 
$R_{\rm HCN/CS}$ (Figures \ref{figure6} and \ref{figure8}). 
Hence, we discuss possible chemical causes 
for enhancing $X_{\rm HCN}$/$X_{\rm HCO^+}$ hereafter. 
Line ratios of buried-AGNs are not discussed 
because they would be (at least partly) affected by IR-pumping. 
However, even focusing on the $X_{\rm HCN}$/$X_{\rm HCO^+}$ only, 
it is still difficult to tightly constrain the various possibilities 
due to insufficient amount of data. 
We hope future high resolution, multi-line, 
and multi-species observations will settle the issue. 

\subsection{XDR, PDR, and high temperature chemistry}\label{sec5.1} 
It has been suggested that the enhanced HCN(1-0) intensity observed in AGNs 
is due to abnormal chemistry realized in XDRs (e.g., \citealt{2005AIPC..783..203K,2008ApJ...677..262K}). 
The molecular abundances and the resultant {\it{column-integrated}} line intensities of our target molecules 
under steady state gas-phase XDRs and PDRs were extensively modeled by \citet{2005A&A...436..397M} and \citet{2007A&A...461..793M}. 
In their XDR models, $R_{\rm HCN/HCO^+}$ can exceed unity only at the surface ($N_{\rm H}$ $\la$ 10$^{22.5}$ cm$^{-2}$) 
of low-to-moderate density gas ($n_{\rm H}$ $\la$ $10^{5}$ cm$^{-3}$), 
where a high X-ray flux (e.g., $F_{\rm X}$ $\ga$ 10 erg s$^{-1}$ cm$^{-2}$) can be expected. 
However, for a larger $N_{\rm H}$ where $F_{\rm X}$ is attenuated, they predicted $R_{\rm HCN/HCO^+}$ $<$ 1. 
This is due to the fact that the range of the ionization rate over which 
$X_{\rm HCO^+}$ is high is much wider than that of HCN under X-ray ionization chemistry (\citealt{1996A&A...306L..21L}). 
Note that one weak point in this model would be that the predicted line intensities are 
considerably low at the region of $N_{\rm H}$ $\la$ 10$^{22.5}$ cm$^{-2}$ (\citealt{2006ApJ...650L.103M,2007A&A...461..793M}), 
which seems to be inconsistent to the prominent HCN(4-3) and HCO$^+$(4-3) emissions observed in AGNs.

Another possible scenario is a high temperature chemistry, 
under which neutral-neutral reactions with 
high reaction barriers are enhanced in general (e.g., \citealt{2010ApJ...721.1570H}). 
This kind of chemistry is efficient in, e.g., hot core-like regions and mechanically dominated regions. 
For example, a formation path of CN + H$_2$ $\rightarrow$ HCN + H has a barrier of 
$\sim$ 960 K (KIDA\footnote{\url{http://kida.obs.u-bordeaux1.fr}}). 
This reaction can be efficient at $T_{\rm kin}$ $\ga$ 300 K (\citealt{2010ApJ...721.1570H}). 
Indeed, such a high temperature can likely be expected in AGNs 
(e.g., \citealt{2008ApJ...677..262K,2012A&A...537A.133D}). 
At that temperature, $X_{\rm HCO^+}$ can be somewhat reduced 
due to an activated reaction with H$_2$O to form H$_3$O$^+$. 
What is important is that this chemistry can be 
complemented to conventional XDR (ionization) models (\citealt{2013ApJ...765..108H}), 
where we can expect much higher gas temperature 
due to efficient X-ray heating than in SB environments (PDR). 
Interestingly, recent high resolution VLA observations revealed a higher fractional abundance of NH$_3$ 
in the very nuclear region of NGC 3079 (a type-2 AGN; \citealt{2015PASJ...67....5M}) 
than in SB galaxies (\citealt{2013A&A...552A..34T}, and references therein). 
NH$_3$ is also a typical molecule efficiently formed in high temperature environments. 
We should note that evaporation from dust grain can also increase 
the $X_{\rm NH_3}$ especially at lower temperature ($\la$ 100 K; \citealt{2001ApJ...546..324R}), 
which will enhance the $X_{\rm HCN}$ via subsequent nitrogen reactions. 
However, at low temperature, ion-neutral reactions are so quick that we can envision, 
e.g., C$^+$ + H$_2$O $\rightarrow$ HCO$^+$ + H, will efficiently enhance the $X_{\rm HCO^+}$ 
(O-bearing species such as H$_2$O should also be ejected from dust). 
This would reduce $X_{\rm HCN}/X_{\rm HCO^+}$, 
which seems to be inconsistent with 
the high $X_{\rm HCN}/X_{\rm HCO^+}$ suggested in this Section \ref{sec4}, 
thus the dust grain reactions will not be very critical. 

Keeping these models in mind, we hereafter present possible interpretation of the $R_{\rm HCN/HCO^+}$ 
in three instructive AGNs and two SB galaxies, namely NGC 7469, NGC 1068, NGC 1097, NGC 253, and M82. 

(i) NGC 7469---We first discuss the $R_{\rm HCN/HCO^+}$ measured at the AGN position of NGC 7469 
(denoted as {\it{N7469 (AGN)}} in Figure \ref{figure1}), 
which is comparable to those of some SB galaxies such as NGC 253 
despite its orders of magnitude higher X-ray luminosity (log $L_{\rm{2-10keV}}$ = 43.2; \citealt{2014ApJ...783..106L}). 
No morphological and kinematic signatures of a jet-ISM interaction has been found in the CND of NGC 7469 
through the high resolution observations of H$_2$ and Br$\gamma$ emission lines 
(\citealt{2009ApJ...696..448H,2011ApJ...739...69M}) 
and then mechanical heating might not play an important role in this galaxy, 
although \citet{2003ApJ...592..804L} found a core jet-like structure at a radio wavelength. 
Regarding the origin of the relatively low $R_{\rm HCN/HCO^+}$, 
\citet{2015ApJ...811...39I} pointed out the compactness of the XDR. 
They estimated the spatial extent of the XDR 
and the region where $F_{\rm X}$ $\ga$ 10 erg s$^{-1}$ cm$^{-2}$ (\citealt{2007A&A...461..793M}) 
to be $\sim$ 42 pc and $\sim$ 35 pc in radius, respectively. 
Therefore, not a poor but a still relatively large observing beam 
employed in the NGC 7469 observations ($\sim$ 150 pc; Table \ref{tbl1}) 
might have picked up line fluxes emanating from the extended SB region, 
which would have resulted in reducing the $R_{\rm HCN/HCO^+}$ of this AGN. 

(ii) NGC 1068---The X-ray luminosity of this AGN 
(log $L_{\rm{2-10keV}}$ = 43.0; \citealt{2012ApJ...748..130M}) 
is quite comparable to that of NGC 7469. 
The degree of contamination from PDRs and SNe 
(i.e., stellar feedback) to the observed line ratio 
is unclear, although we expect it is relatively low 
by considering the moderate stellar age inside the CND 
(200-300 Myr; \citealt{2007ApJ...671.1388D}). 
This age is much older than those of SB galaxies, 
e.g., NGC 253 ($\sim$ 6 Myr; \citealt{2009MNRAS.392L..16F}) 
and M82 ($\sim$ 10 Myr; \citealt{2003ApJ...599..193F}). 
We found a relatively high $R_{\rm HCN/HCO^+}$ of $\sim$ 1.5 at its precise AGN position 
(denoted as {\it{N1068 (AGN)}} in Figure \ref{figure1}), 
which value is well consistent with the results in \citet{2014A&A...567A.125G}. 
The ratio was measured with a single synthesized beam of 35 pc. 
\citet{2014A&A...567A.125G} also reported a correlation between the spatially resolved 
HCN(4-3) to CO(3-2) integrated intensity ratio ($\equiv$ $R_{\rm HCN/CO}$) 
and the 6-8 keV X-ray flux ($\equiv$ $R_{\rm X-ray/CO}$) across the CND of NGC 1068. 
The same trend can also be found in HCO$^+$(4-3), SiO(2-1), and CN(2-1) 
(\citealt{2010A&A...519A...2G,2014A&A...567A.125G}), which suggests that the whole CND is a giant XDR. 
However, again according to \citet{2007A&A...461..793M}, 
to reproduce $R_{\rm HCN/HCO^+}$ $>$ 1 at this AGN position 
is quite difficult since the $N_{\rm H_2}$ toward the AGN position 
expected from the results of the CO multi-line non-LTE analysis by \citet{2014A&A...570A..28V} 
is $\sim$ 6 $\times$ 10$^{23}$ cm$^{-2}$ (see the footnote of Section \ref{sec2} for the assumptions in this estimation). 
To interpret the high $R_{\rm HCN/HCO^+}$ of {\it{N1068 (AGN)}} is thus not straightforward at all. 
We might have to add some other mechanisms such as high temperature chemistry 
(\citealt{2010ApJ...721.1570H,2013ApJ...765..108H}) to forcibly enhance $X_{\rm HCN}$. 

As for the $R_{\rm HCN/HCO^+}$ measured at other positions, 
e.g., the East/West-knots of the CND of NGC 1068 
(namely {\it{N1068 (E-knot)}} and {\it{N1068 (W-knot)}} in Figure \ref{figure1}), 
we found they are 2-times higher than that of {\it{N1068 (AGN)}}. 
This is a bit surprising result since the ratio 
increases as receding from the nucleus (= the X-ray peak). 
We speculate the cause of this enhancement at the E/W-knots is likely related to 
that for the prominent 2.12 $\mu$m H$_2$ emission there (e.g., \citealt{2009ApJ...691..749M}). 
The $R_{\rm HCN/CO}$-$R_{\rm X-ray/CO}$ correlation shown above 
and the very uniform distribution of the H$_2$ $\lambda$2.25 $\mu$m/$\lambda$2.12 $\mu$m ratio 
across the CND (\citealt{2014MNRAS.442..656R}) seem support 
a scenario that these emissions are primarily 
due to (non-directional) X-ray heating (see also \citealt{2003A&A...412..615G}). 
Note that the H$_2$ line ratio at the CND ($\sim$ 0.1) is 
a typical value for thermal processes by 
X-ray and/or shock heating (\citealt{1994ApJ...427..777M}). 
In this scenario, we might be observing HCN and HCO$^+$ 
emissions emerged from molecular clouds directly illuminated by X-ray radiation 
(i.e., clouds directly seen by the AGN itself, not strongly intercepted by a dusty torus) at the East/West-knots, 
considering the geometry of the type-2 nucleus 
(the position angle of the almost edge-on H$_2$O maser disk is -45$^\circ$; \citealt{1996ApJ...472L..21G}) 
and the surrounding CND (the inclination angle is $\sim$ 41$^\circ$; \citealt{2014A&A...567A.125G}). 
If so, the X-ray fluxes received at the East/West-knots are quite high because of little attenuation, 
which can result in efficient X-ray heating and thus high $X_{\rm HCN}/X_{\rm HCO^+}$ 
due to, e.g., high temperature neutral-neutral reactions (\citealt{2010ApJ...721.1570H,2013ApJ...765..108H}). 
Indeed, if there is no attenuation, we can expect $F_{\rm X}$ 
is as high as $\sim$ 100 erg s$^{-1}$ cm$^{-2}$ even at 30 pc away from the nucleus. 
According to the model calculation by \citet{2007A&A...461..793M}, this high X-ray flux can heat up the gas 
with $n_{\rm H_2}$ = 10$^{5-6}$ cm$^{-3}$ to $\gg$ 100 K. 

On the other hand, the radial expansion of the CND of NGC 1068 
(\citealt{2011ApJ...736...37K,2014MNRAS.445.2353B}) would support the existence of 
shocks and/or a jet-ISM (and possibly, outflow-ISM) interaction, 
which can also contribute to excite the H$_2$ molecule. 
We especially mention that \citet{2014A&A...567A.125G} suggested 
that as much as $\sim$ 50\% of CO(3-2) emission 
stems from the outflowing component, 
which extends out to $\sim$ 400 pc away from the nucleus. 
Hence, there is at least a certain portion of a region dominated by mechanical heating 
(MDR; e.g., \citealt{2011A&A...525A.119M,2012A&A...542A..65K,2015A&A...574A.127K}) in the CND. 
Again high gas temperature chemistry would be responsible for the high $R_{\rm HCN/HCO^+}$ in this case. 
Indeed, we can expect the maximum temperature of $n_{\rm H_2}$ = 10$^5$ cm$^{-3}$ gas 
should reach as high as $\sim$ 2500 K for C-shock 
when the shock velocity is $>$ 20 km s$^{-1}$ (\citealt{2013MNRAS.436.2143F}), 
which is high enough to produce prominent H$_2$ emission. 
Such a high temperature is well beyond the reaction barrier of 
HCN formation from CN, thus HCN can be abnormally abundant. 
According to the model calculation of molecular abundances 
in MDRs by \citet{2012A&A...542A..65K}, 
$X_{\rm HCN}$/$X_{\rm HCO^+}$ can well exceed 100 
when the input mechanical energy is $\ga$ 10$^{-18}$ erg s$^{-1}$ cm$^{-3}$, 
which is large enough to produce high $R_{\rm HCN/HCO^+}$. 
Interestingly, the $R_{\rm HCN/CO}$-$R_{\rm X-ray/CO}$ correlation in NGC 1068 
has large scatter at lower $R_{\rm X-ray/CO}$, 
i.e., away from the AGN (\citealt{2014A&A...567A.125G}), 
suggesting the importance of other heating mechanisms 
than X-ray radiation for driving the underlying chemistry. 
Indeed, the line velocity dispersion is high at the East/West-knots (\citealt{2014A&A...567A.125G}). 
In this scenario, it would also be probable that the outflow/shock velocity is so high at {\it{N1068 (AGN)}} 
that the shock wave provides ionized gas to increase ions 
(e.g., \citealt{1980ApJ...237L..43D,1983ApJ...267..174E,1997ApJ...485..263K,2004MNRAS.351.1054R}), 
which would potentially result in a depressed $X_{\rm HCN}/X_{\rm HCO^+}$ near the nucleus. 

(iii) NGC 1097---Another remarkable example is 
the $R_{\rm HCN/HCO^+}$ measured at the AGN position of 
NGC 1097 ({\it{N1097 (AGN)}} in Figure \ref{figure1}). 
As we stated in Section \ref{sec4.5}, we would need enhanced 
$X_{\rm HCN}$/$X_{\rm HCO^+}$ compared to SB galaxies 
to yield this high $R_{\rm HCN/HCO^+}$ 
when $n_{\rm H_2}$ $\sim$ 10$^5$ cm$^{-3}$ (\citetalias{2013PASJ...65..100I}). 
However, judging from the very low X-ray luminosity 
(log $L_{\rm 2-10keV}$ = 40.8; \citealt{2014ApJ...783..106L}), 
the XDR chemistry seems not to play 
an important role in NGC 1097 (\citetalias{2013PASJ...65..100I}). 
Rather, considering the current spatial resolution (94 pc) for the observations (Table \ref{tbl1}), 
there is another possibility that the line emitting region is very close to the AGN where we can still expect 
a high X-ray flux, or a high energy deposition rate per particle (e.g., \citealt{1996ApJ...466..561M}). 
Indeed, the line luminosity of HCN(4-3) in NGC 1097 is totally comparable to 
that of the individual molecular clump seen at the center of NGC 253 ($\sim$ 20 pc in size; \citealt{2011ApJ...735...19S}), 
suggesting the similarly small size of the line emitting region in NGC 1097. 
Radiative feedback from SB activity would be discarded 
considering the high gas temperature ($\ga$ 100 K) suggested for 
the molecular phase (\citetalias{2013PASJ...65..100I}; \citealt{2012ApJ...751..144B}). 
On the other hand, mechanical heating can be another choice, 
which is supported by the detection of a compact radio jet 
(largest beam-convolved and projected size is $\sim$ 90 pc; \citealt{2000MNRAS.314..573T}). 
A tentative gradient of HCN(1-0)/HCO$^+$(1-0) line ratio inside the CND 
(lower towards the center, and higher towards the outer edge of the CND; \citealt{2015A&A...573A.116M}) 
would also support this scenario, although \citet{2015A&A...573A.116M} did not fully resolve the CND. 
Note that an inferred SNe rate is quite low as $O(10^{-4})$ yr$^{-1}$ (\citealt{2007ApJ...671.1388D}), 
thus we can neglect its influence on the chemistry. 

(iv) NGC 253 and M82---As for the SB galaxies, 
the predicted $R_{\rm HCN/HCO^+}$ in \citet{2007A&A...461..793M} is almost consistent with, 
but seems to be a factor of a few higher than the observed values. 
This can be explained by an enhanced cosmic ray ionization rate due to frequent supernovae (SNe), 
which would increase $X_{\rm HCO^+}$ even in a dense molecular cloud 
(e.g., \citealt{2011MNRAS.414.1583B,2011A&A...525A.119M,2013A&A...549A..39A}). 
If this is true, $R_{\rm HCN/HCO^+}$ can reflect the evolutionary phase of SB activity. 
The lower $R_{\rm HCN/HCO^+}$ in M82 than in NGC 253 follows 
this scenario well (\citetalias{2013PASJ...65..100I}; \citealt{2008ApJ...677..262K}), 
although we might be able to construct some counter arguments as well 
(\citealt{2007ApJ...656..792P,2014ApJ...784L..31Z}). 
We mention that mechanical heating seems to be responsible 
for the relatively high $R_{\rm HCN/HCO^+}$ in young SB galaxy NGC 253 
(SB age $\sim$ 6 Myr; \citealt{2009MNRAS.392L..16F}), 
whose ratio is totally comparable to those of {\it{N7469 (AGN)}} and {\it{N1068 (AGN)}}. 
Indeed, this heating mechanism is claimed to be important in NGC 253 (\citealt{2014A&A...564A.126R}). 

\subsection{Effects of metallicity and elemental abundance}\label{sec5.2}
The chemistry of PDRs in high metallicity (i.e., Z $\geq$ Z$_\odot$ where Z$_\odot$ denotes the solar metallicity) 
environments was modeled by, e.g., \citet{2012MNRAS.424.2646B}. 
They showed $X_{\rm HCN}$ increases with metallicity, 
whereas $X_{\rm HCO^+}$ and $X_{\rm CS}$ are insensitive to the variation, 
when FUV radiation field and cosmic ray ionization (mimicking X-ray ionization) rate are fixed. 
From the observational side, \citet{2013MNRAS.433.1659D} found a positive correlation between gas-phase metallicity 
and $N_{\rm HCN}/N_{\rm CS}$ in the sample of early type galaxies, metal-rich spirals, and SB galaxies, 
which supported the Bayet et al.'s prediction. 
However, if we regard 12 + $\log$ [O/H] as an indicator of the gas phase metallicity, 
NGC 1068, NGC 1097, NGC 253, and M82 show almost the same value (\citealt{2008ApJ...672..214G}, and references therein). 
Thus, we suggest that the metallicity is not an important factor 
for the submm-HCN diagram at least when we treat already chemically evolved systems 
such as the central regions of galaxies studied in this work. 

On the other hand, it seems likely that an overabundance of elemental nitrogen 
observed in AGNs (e.g., \citealt{1991MNRAS.249..404S}) can contribute to the HCN-enhancement, 
since $X_{\rm HCN}$ depends on the elemental abundance of nitrogen (\citealt{2008ApJ...676..978B}). 
Interestingly, the HCN(1-0)/HCO$^+$(1-0) line ratio is $\sim$ 4-times lower 
in the LMC than in nearby massive SB galaxies (\citealt{2014ApJ...793...37A}). 
Although this can be partly reconciled by the different gas density in each galaxy, 
we can also expect $\sim$ 2-3 times lower N/O elemental abundance ratio in the LMC than in 
the solar metallicity Galactic environment (\citealt{2009A&A...496..841H}) 
would have influenced the $X_{\rm HCN}/X_{\rm HCO^+}$. 

\subsection{IR-pumping}\label{sec5.3} 
This mechanism must be vital for vibrationally excited lines of 
these dense gas tracers such as HCN($v_2$=1$^{1f}$, $J$=4-3), 
since their energy levels ($\sim$ 1000 K) are too high to be collisionally excited. 
Although the limited extragalactic detections of the vibrationally excited lines 
(\citealt{2010ApJ...725L.228S,2013AJ....146...91I,2015A&A...574A..85A}) limits our discussion on this topic, 
it would be at least influencing the rotational population and thus the line intensity of HCN in NGC 4418 (\citealt{2010ApJ...725L.228S}). 
Meanwhile, HCN($v_2$=1$^{1f}$, $J$=4-3) emission was not detected in NGC 7469 (\citealt{2015ApJ...811...39I}) 
in spite of its high IR luminosity of $L_{\rm 8-1000 \mu m}$ = 10$^{11.4}$ $L_\odot$, 
which is comparable to that of NGC 4418 (\citealt{2003AJ....126.1607S}). 
This would be due to the optical depth effect; 
we can not detect a fully optically thin emission even if it surely exists. 
The column density of HCN($v_2$=1$^{1f}$, $J$=4-3) would not be so large along the line of sight 
towards the CND of NGC 7469 that the emission is optically thin. 
We speculate that the vibrationally excited lines can only be visible in compact (i.e., close to a warm IR-source), 
heavily obscured (i.e., high column density) nuclei such as in NGC 4418, 
where we can expect HCN($v_2$=1$^{1f}$, $J$=4-3) line emission is (moderately) optically thick. 

\subsection{Time-dependent chemistry}\label{sec5.4}
This can be important since the $X_{\rm HCN}/X_{\rm HCO^+}$ is predicted to be 
highly time-dependent with possible variations of orders of magnitude 
especially when $\geq$ 10$^4$ yr has passed 
from the ignition of chemical reactions (\citealt{2013JPCA..117.9593M}). 
This mechanism would influence the chemistry more in nuclear regions of galaxies than in quiescent regions, 
since the former region is more dynamic than the latter. 
The time-evolution is typically studied via time clock molecules such as sulfur bearing species. 
Indeed, while $N_{\rm CS}/N_{\rm SO}$ in the CNDs of NGC 1097 and NGC 1068 estimated in LTE analysis 
are $\sim$ 2-3 (\citealt{2015A&A...573A.116M,2015PASJ...67....8N}), 
it is just $\sim$ 0.2 in the Orion hot core (Table 2 in \citealt{2014A&A...567A..95E}). 
If we regard the above CNDs are influenced by mechanical heating, 
and the abundances of these species in NGC 1097 and NGC 1068 
depend on the evolution of hot core-like systems, 
this result indicates that NGC 1097 and NGC 1068 are more evolved systems 
than Orion hot core (e.g., \citealt{2014A&A...567A..95E,1997ApJ...481..396C}). 
Future sensitive and systematic study of time clock species will help us understand the chemical evolution in AGNs.

\section{Summary and conclusion}\label{sec6}
In this paper, we aim at up-dating the submm-HCN diagram 
which was tentatively proposed by \citetalias{2013PASJ...65..100I} 
and investigating the potential origin of the enhanced HCN(4-3) intensity 
with respect to HCO$^+$(4-3) and CS(7-6) in AGNs (i.e., HCN-enhancement). 
The main results and conclusions are summarized as follows:

\begin{itemize}
\item[-] Compiling data from the literature and the ALMA archive, 
we up-dated the submm-HCN diagram (Figure \ref{figure1}), 
first proposed by \citetalias{2013PASJ...65..100I}. 
The number of the data point is significantly increased by $\sim$ 5 times (Figure \ref{figure2}). 
As was supposed by \citetalias{2013PASJ...65..100I}, 
we found a clear trend that AGNs tend to show 
higher $R_{\rm HCN/HCO^+}$ and/or $R_{\rm HCN/CS}$ than SB galaxies 
when these samples are observed at high resolutions enough to separate 
regions energetically dominated by AGNs (e.g., XDR) 
from those contaminated by co-existing SB activities. 
When AGNs are observed at low resolutions ($\ga$ 1 kpc), 
the energetics within the beam seems to be 
dominated by surrounding SB activities, which would result in reduced line ratios. 
\item[-] Simple non-LTE radiative transfer modelings 
involving both collisional and radiative excitation were conducted with the RADEX code. 
Under the constant density of $n_{\rm H_2}$ = 10$^5$ cm$^{-3}$, 
we found $R_{\rm HCN/HCO^+}$ is not so sensitive to excitation, 
whereas $R_{\rm HCN/CS}$ strongly depends on it, 
especially when CS(7-6) is optically thin and HCN(4-3) is thick. 
\item[-] From the above non-LTE modelings, we suggest that both 
$X_{\rm HCN}/X_{\rm HCO^+}$ and $X_{\rm HCN}/X_{\rm CS}$ 
would be significantly enhanced in AGNs by several to even $\ga$ 10 times 
than those in SB galaxies, to reproduce their line ratios. 
From this perspective, we suggest that 
the variation of molecular abundances would 
drive the HCN-enhancement in AGNs. 
\item [-] On the other hand, we could not fully discard the possibility 
that the systematically higher gas density in AGNs than in SB galaxies 
is a cause of the HCN-enhancement, 
which was revealed by varying $n_{\rm H_2}$ to a higher value (here 5 $\times$ 10$^6$ cm$^{-3}$), 
although the feasibility of such a difference is unclear within 
the central $\sim$ 100 pc regions of galaxies. 
\item[-] We also investigated HCN/HCO$^+$ and HCN/CS 
line ratios at other transitions based on our RADEX models. 
As a result, we found enhanced $X_{\rm HCN}/X_{\rm HCO^+}$ 
and $X_{\rm HCN}/X_{\rm CS}$ subsequently 
result in high line ratios at any transition, 
that seems to be consistent with observations. 
\item[-] Various possible chemical scenarios for 
the high $X_{\rm HCN}/X_{\rm HCO^+}$ in AGNs are discussed. 
Although it is still far from being well understood, 
we suppose it would be difficult to explain the high line ratios in AGNs 
solely by conventional X-ray ionization models. 
We conclude that some additional mechanisms, 
e.g., high temperature chemistry (likely related to mechanical heating), 
seem to be necessary to fully explain the observations. 
\end{itemize}

In this work, we preferred the scenario of enhanced HCN abundance with respect to HCO$^+$ and CS, 
rather than a systematically higher gas density in AGNs than SB galaxies (Section \ref{sec4.4}). 
However, the latter scenario was not fully discarded because of the simplified modelings in this work. 
Indeed, the observed line ratios are obtainable using a wide parameter space of physical and chemical conditions. 
To fully solve this problem, we have to conduct multi-line, multi-species non-LTE modelings 
of both AGNs and SB galaxies with sufficiently high resolutions to spatially resolve their CNDs, 
as was recently conducted by \citet{2014A&A...570A..28V}. 
Moreover, as presented in Section \ref{sec6}, 
there are still various competing chemical interpretations for the HCN-enhancement. 
Higher resolution observations of not only typical and relatively 
bright dense molecular gas tracers (e.g., HCN, HCO$^+$, HNC, and CS), 
but also rarer species which are more susceptible to the variation of the underlying chemistry 
(e.g., NH$_3$, CH$_3$CN, HNCO, CH$_3$OH, and SiO) will be inevitable to test them.

\acknowledgments
We appreciate the anonymous referee for the careful reading 
and for the very kind comments to improve this paper. 
This paper makes use of the following ALMA data: ADS/JAO.ALMA\#2011.0.00083.S. 
ALMA is a partnership of ESO (representing its member states), NSF (USA), and NINS (Japan), 
together with NRC (Canada) and NSC and ASIAA (Taiwan), in cooperation with the Republic of Chile. 
The Joint ALMA Observatory is operated by ESO, AUI/NRAO, and NAOJ. 
The National Radio Astronomy Observatory is a facility of the National Science Foundation 
operated under cooperative agreement by Associated Universities, Inc. 
In addition, this research has made use of the NASA/IPAC Extragalactic Database (NED) 
which is operated by the Jet Propulsion Laboratory, California Institute of Technology, 
under contract with the National Aeronautics and Space Administration.
T. I. was supported by the ALMA Japan Research Grant of NAOJ Chile Observatory, NAOJ-ALMA-0029 and -0075. 
T. I. is thankful for the fellowship received from 
the Japan Society for the Promotion of Science (JSPS).






\bibliography{reference}

\appendix
\section{A. Notes on individual galaxies}\label{app-A}
This Appendix provides some basic information on the individual galaxies collected 
in Table \ref{tbl1} and Figure \ref{figure1}, reported in previous studies. 
The information was compiled mainly from the perspective of nuclear energetics as mentioned in Section \ref{sec2}. 

\subsection{NGC 1068}
Undoubtedly this is one of the best studied Seyfert galaxies hosting a luminous type-2 AGN 
($L_{\rm{2-10keV}}$ = 10$^{43.0}$ erg s$^{-1}$; \citealt{2012ApJ...748..130M}). 
Broad Balmer emission lines are observed in the polarized light (\citealt{1985ApJ...297..621A}), 
which is a supportive evidence for the unified model of AGNs (\citealt{1993ARA&A..31..473A}). 
In X-rays, this AGN is known to be substantially Compton thick with 
$N_{\rm H}$ $\sim$ 10$^{25}$ cm$^{-2}$ along the line of sight (e.g., \citealt{2000MNRAS.318..173M,2012ApJ...748..130M,2015arXiv151103503M}). 
NGC 1068 also hosts a ring- or arm-like SB region with a diameter of $\sim$ 30$''$ (\citealt{1988ApJ...334..573T}). 
Several unbiased line surveys with single dish telescopes and/or interferometric observations of 
individual molecular line emissions at millimeter/submillimeter 
have been conducted towards this AGN (e.g., \citealt{2004A&A...419..897U,2011ApJ...736...37K,2013A&A...549A..39A,
2014PASJ...66...75T,2014A&A...567A.125G,2015PASJ...67....8N}). 
The CND is bright not only in CO but also, and more distinctively, in dense gas tracers such as HCN and HCO$^+$ lines 
(e.g., \citealt{2000ApJ...533..850S,2008Ap&SS.313..279K,2014A&A...567A.125G}). 
This CND mainly consists of two bright knots, namely, East and West knots, which are $\sim$ 100 pc away from the exact location of the AGN. 
Interestingly, HCN(4-3)/HCO$^+$(4-3) line ratio is two times higher at these knots than at the AGN location (\citealt{2014A&A...567A.125G}). 
NGC 1068 is also known to possess a radio jet, ionized outflow, and prominent molecular outflow, 
which might be interacting with the CND (e.g., \citealt{2011ApJ...736...37K}). 

\subsection{NGC 1097}
This galaxy hosts a low luminosity type-1 AGN (LLAGN, $L_{\rm{2-10keV}}$ = 10$^{40.8}$ erg s$^{-1}$; \citealt{2014ApJ...783..106L}) 
as evidenced by double-peaked broad Balmer emission lines with time variability (FWHM $\sim$ 7500 km s$^{-1}$; \citealt{1997ApJ...489...87S}). 
Interestingly, NGC 1097 also showed a transient phenomenon from a LINER nucleus to a type-1 AGN. 
This LLAGN and the CND are surrounded by a circumnuclear SB ring with a radius of $\sim$ 10$''$ = 700 pc (\citealt{1995AJ....110.1009B}). 
At infrared, \citet{2007ApJ...659..241M} reported the absence of 3.3 $\mu$m PAH feature at sub-arcsec scale in the CND, 
which is typical for strong radiation field. 
But attempts to fit the IR-data with clumpy torus models failed, 
which led \citet{2007ApJ...659..241M} to conclude that the torus is absent or weak. 
Alternatively, they argued a nuclear star-forming cluster, 
which is likely co-existing with the LLAGN (\citealt{2005ApJ...624L..13S}) would be dominant at least in MIR energetics. 
Both the CND and the SB ring accompany large amount of molecular gas 
prominent in e.g., CO, HCN, HCO$^+$ emissions (e.g., \citetalias{2013PASJ...65..100I};\citealt{2008ApJ...683...70H,2012ApJ...747...90H,2003PASJ...55L...1K}). 
It has been claimed that the CND is warm even in the molecular phase ($\sim$ several hundreds K; \citetalias{2013PASJ...65..100I}; \citealt{2012ApJ...751..144B}),  
which would be heated by the AGN through radiatively and/or mechanically. 

\subsection{NGC 1365}
This archetypal barred spiral galaxy hosts both a Seyfert 1.5 nucleus 
($L_{\rm{2-10keV}}$ = 10$^{42.3}$ erg s$^{-1}$; \citealt{2014ApJ...783..106L}) with time variability (\citealt{2014ApJ...788...76W}) 
and a circumnuclear SB ring with a radius of 5$''$-10$''$. 
Several star clusters are embedded inside the SB ring (\citealt{2005A&A...438..803G}). 
A comprehensive review of this galaxy can be found in \citet{1999A&ARv...9..221L}. 
\citet{2012MNRAS.425..311A} estimated that the AGN contributes only $\sim$ 5\% of the total IR emission 
in the central $\sim$ 5 kpc, although it seems to dominate the energetics at $\lambda$ $\la$ 24 $\mu$m. 
Previous CO multi-transition observations revealed that most of their emissions comes from the SB ring, 
and little is from around the Seyfert nucleus (\citealt{2007ApJ...654..782S}). 
This suggests that the primary heating source of molecular gas is the SB activity in the ring, rather than the AGN, 
although the spatial distribution of dense gas tracers such as HCN and HCO$^+$ is currently unknown. 
High resolution observations of them are thus desirable. 

\subsection{NGC 4945}
This galaxy contains an AGN as evidenced by its strong and rapidly variable X-ray emission (\citealt{2000A&A...356..463G,2014ApJ...793...26P}). 
The intrinsic 2-10 keV luminosity is estimated to be $L_{\rm{2-10keV}}$ = 10$^{42.5}$ erg s$^{-1}$ (\citealt{2000A&A...356..463G}). 
The AGN is classified as Seyfert 2 with a huge amount of obscuring material ($N_{\rm{H}}$ = 10$^{24.5}$ cm$^{-2}$; \citealt{2014ApJ...793...26P}). 
It has also been found that NGC 4945 exhibits a prominent SB activity including a large IR luminosity and IRAS color similar to a giant HII region, 
but almost an edge-on view ($i$ = 78$^\circ$) has prevented detailed study at optical wavelengths. 
\citet{2000A&A...357...24M} reported that the spatial distribution of the star bursting region is ring-like with a radius of $\sim$ 2.5$''$ (50 pc), 
which is visible most clearly in a Pa$\alpha$ map. 
\citet{2007ApJ...670..116C} argued that this SB activity, rather than the AGN, is the dominant heating source for dust in the nuclear region. 
Furthermore, interferometric HCN(1-0), HCO$^+$(1-0), and HNC(1-0) observations revealed that 
these dense gas tracers are emitted from an inclined rotating disk-like component with a radius of $\sim$ 4$''$ (\citealt{2005MNRAS.364...37C}), 
i.e., emanating from both around the AGN and the SB ring. 
Higher resolution observations are thus necessary to obtain reliable spectrum of each component.

\subsection{NGC 7469}
The existence of a type-1 AGN ($L_{\rm{2-10keV}}$ = 10$^{43.2}$ erg s$^{-1}$; \citealt{2014ApJ...783..106L}) in this galaxy is evidenced 
by broad Balmer emission lines (FWHM $\sim$ 4370 km s$^{-1}$; \citealt{2014ApJ...795..149P}) 
with time variability (\citealt{1990ApJ...353..445B,1998ApJ...500..162C}). 
Time variability is also confirmed by UV and X-ray observations (e.g., \citealt{2000ApJ...544..734N,2005ApJ...634..193S}). 
A core jet-like structure (\citealt{2003ApJ...592..804L}) and ionized outflows (\citealt{2007A&A...466..107B}) 
have been found, although high resolution H$_2$ observations 
could not detect any signature of jet-ISM (or outflow-ISM) interaction in the CND (e.g., \citealt{2009ApJ...696..448H}). 
Similar to some other Seyfert galaxies, the AGN and the CND of NGC 7469 are surrounded by a prominent SB ring 
with a radius of 1.5$''$-2.5$''$ (\citealt{2003AJ....126..143S,2007ApJ...661..149D}), 
which accounts for two-thirds of the bolometric luminosity of the galaxy (\citealt{1995ApJ...444..129G}). 
Inside the nuclear $r$ $\la$ 130 pc region, \citet{2007ApJ...671.1388D} reported an existence of star clusters with modest ages of 110-190 Myr. 
As for the molecular phase, \citet{2015ApJ...811...39I} reported the significant 
concentration of HCN(4-3), HCO$^+$(4-3), and CS(7-6) line emissions towards the CND.  

\subsection{NGC 4418}
This is a luminous infrared galaxy (LIRG) with an infrared luminosity of 
$L_{\rm 8-1000 \mu m}$ $\sim$ 10$^{11}$ $L_\odot$ (\citealt{2004AJ....128.2037I}). 
NGC 4418 shows no clear AGN signature at optical and X-ray (\citealt{1989ApJ...347..727A,2003MNRAS.344L..59M}), 
However, the 5-23 $\mu$m MIR spectrum shows a typical feature of an obscured AGN (\citealt{2001A&A...365L.353S}). 
Furthermore, \citet{2004AJ....128.2037I} found that the nuclear star forming activity estimated from PAH emission 
can account for only 1/50 of the total IR luminosity, suggesting the primary energy source is deeply embedded in dust and gas. 
Indeed, \citet{2013ApJ...764...42S} estimated the dust continuum opacity at 860 $\mu$m is surprisingly as large as $\tau_{\rm 860 \mu m}$ $\sim$ 1, 
suggesting a concordantly large hydrogen column density of $N_{\rm H}$ $\sim$ 10$^{26}$ cm$^{-2}$. 
As for the hidden energy source, \citet{2004AJ....128.2037I} also suggested that the observed large equivalent width of H$_2$ emission 
and a high HCN(1-0)/HCO$^+$(1-0) ratio at the nucleus can be well explained by the presence of a strong X-ray source such as an AGN. 
Note that shock heating due to in/outflows detected by molecular observations 
(e.g., \citealt{2012A&A...541A...4G,2013ApJ...764...42S}) would also influence these features. 
Considering the above, we classify this galaxy as a buried-AGN in this paper, although the evidence is still tentative. 

\subsection{IRAS12127-1412}
This is a ULIRG with $L_{\rm 8-1000 \mu m}$ = 10$^{12.1}$ $L_{\odot}$ (\citealt{2006ApJ...637..114I}), 
consists of two nuclei with a separation of $\sim$ 10$''$ (\citealt{2002ApJS..143..277K}). 
Optical spectroscopy classified this object as a non-Seyfert (LINER or HII region; \citealt{1999ApJ...522..113V}), 
but the nuclear source is obscured as indicated by a large silicate optical depth of $\sim$ 2.5-3.0 (\citealt{2007ApJS..171...72I}). 
IR observations, on the other hand, suggest that the energy source of the North-East nucleus is very compact, 
which is further characterized by weak PAH emissions and a steep temperature gradient (\citealt{2006ApJ...637..114I,2007ApJS..171...72I}). 
These are characteristic features to buried-AGNs, thus we classify this object as such. 
Note that IRAS12127-1412 has a 1.4 GHz radio luminosity which is $\sim$ 4 times 
brighter than the value expected for a star forming galaxy (\citealt{2009ApJ...702L..42S}), suggesting the existence of an AGN-jet. 
This hypothesized jet may explain the observed outflow signatures in [NeIII] line (\citealt{2009ApJ...702L..42S}) as well. 

\subsection{M82}
This galaxy, as well as NGC 253, is one of the most studied galaxies after Milky Way. 
It is famous for its intense star formation (SFR $\sim$ 10 $M_\odot$ yr$^{-1}$; \citealt{2003ApJ...599..193F}) 
and a prominent superwind at many wavelengths (e.g., \citealt{1995ApJ...439..155B,2002PASJ...54..891O}). 
The current SB region extends over $\sim$ 500 pc centered on the nucleus ({\it{SB core}}), 
which consists of four high surface brightness clumps denoted as A, C, D, and E in \citet{1978ApJ...221...62O}. 
The bolometric luminosity of the SB core is $\sim$ 10$^{10.8}$ $L_{\odot}$, half of which is emanating from OB stars (\citealt{2003ApJ...599..193F}). 
Higher resolution HST observations revealed that $\sim$ 200 super star clusters exist in that nuclear region (e.g., \citealt{2005ApJ...619..270M}). 
The SB history was extensively modeled by \citet{1993ApJ...412...99R} and \citet{2003ApJ...599..193F}, 
who predicted that the bursts occurred $\sim$ 10 and 5 Myr ago each lasting for a few Myr. 
The estimated age is consistent to the observed molecular properties 
suggesting M82 hosts an evolved SB whose energetics is dominated 
by radiative processes (i.e., PDRs) rather than shocks (e.g., \citealt{2011A&A...525A..89A}). 
Note that it has been postulated that M82 hosts an intermediate mass black hole 
located at $\sim$ 600 pc away from the nucleus (e.g., \citealt{2006MNRAS.370L...6P,2014Natur.513...74P}). 

\subsection{NGC 253}
The star forming region of this nearest and IR-brightest SB galaxy is located in the central $r$ $\la$ 500 pc (e.g., \citealt{1985ApJ...289..129S}). 
High resolution millimeter/submillimeter observations revealed that molecular gas in the nuclear region 
is distributed along a 700 $\times$ 200 pc bar-like structure containing several dense clumps 
(e.g., \citealt{1995ApJ...454L.117P,2006ApJ...636..685S,2011ApJ...735...19S}). 
The SB activity has been taking place for at least a few Myr, 
producing an IR luminosity of $\sim$ 10$^{10.5}$ $L_\odot$ (\citealt{1980ApJ...235..392T}), 
or a SFR of $\sim$ 5 $M_\odot$ yr$^{-1}$ assuming a Kennicutt-Schmidt relation (\citealt{1998ApJ...498..541K}). 
The nuclear region consists of a number of massive stellar clusters (e.g., \citealt{2009MNRAS.392L..16F}), 
some of which have a radio counterpart (\citealt{1997ApJ...488..621U}). 
Assuming an instantaneous SB, the age of the stellar clusters is estimated to be $\sim$ 6 Myr, 
whereas they would consist of young ($\la$ 4 Myr) and old ($\sim$ 10 Myr) populations in actual (\citealt{2009MNRAS.392L..16F}). 
Prominent ionized and molecular outflows are emerging from this region as well (e.g., \citealt{2002ApJ...576L..19W,2013Natur.499..450B}). 
Considering the relatively young stellar age, NGC 253 would be in an early phase of a SB evolution. 
Indeed, \citet{2006ApJS..164..450M} suggested that the low-velocity shocks rather than UV radiation would dominate the molecular chemistry in NGC 253. 

\subsection{NGC 1614}
The merging system NGC 1614 is classified as a LIRG ($L_{\rm 8-1000 \mu m}$ = 10$^{11.6}$ $L_\odot$; \citealt{2013AJ....146...47I}), 
which has a bright optical center and two spiral arms at scales of a few kpc. 
At optical wavelengths, it is classified as a SB galaxy (\citealt{1995ApJS...98..171V,2010A&A...518A..10V}). 
Indeed, 2.5-30 $\mu$m IR spectrum, prominent 3.3 $\mu$m PAH emission and 4.05 $\mu$m Br$\alpha$ emission (and their ratios to IR luminosity), 
2.3 $\mu$m stellar CO absorption feature, are typical of a SB galaxy (\citealt{2001ApJ...546..952A,2006ApJ...653.1129B,2010ApJ...721.1233I}), 
and there is no clear indication of an AGN activity. 
The star forming activity is significantly concentrated towards the nuclear region as indicated by 
a SB ring with a radius of $\sim$ 300 pc clearly visible 
at Pa$\alpha$ emission (\citealt{2001ApJ...546..952A}) and radio emission (e.g., \citealt{2010A&A...513A..11O}). 
A low HCN(1-0)/HCO$^+$(1-0) ratio is also supportive of dominance of SB activity (\citealt{2011A&A...528A..30C}). 

\subsection{NGC 3256}
The late-stage merging system NGC 3256 is the most luminous galaxy within 40 Mpc 
with $L_{\rm 8-1000 \mu m}$ = 10$^{11.6}$ $L_\odot$ (\citealt{2003AJ....126.1607S}), 
which has a double nucleus with a separation of $\sim$ 5$''$ = 850 pc. 
The existence of the double nucleus has been revealed at radio, NIR, X-ray, 
as well as molecular line emission (\citealt{1995ApJ...446..594N,1996A&A...305..107K,2002MNRAS.330..259L,2006ApJ...644..862S}). 
Although NGC 3256 is luminous in X-ray ($L_{\rm 0.5-10 keV}$ $\sim$ 10$^{42}$ erg s$^{-1}$; \citealt{2002MNRAS.330..259L}), 
the presence of an AGN is not supported by a considerable amount of observational evidence 
(\citealt{2000AJ....120..645L,2002MNRAS.330..259L,2004MNRAS.352.1335J}). 
The relatively strong X-ray emission would be stellar origin 
since hundreds of young stellar clusters have been found 
in the central region of this system (\citealt{1999AJ....118..752Z,2002AJ....124..166A}). 
Indeed, the above mentioned X-ray luminosity is consistent with 
the boundary to separate AGNs from SB galaxies found in deep X-ray surveys (e.g., \citealt{2005ApJ...632..736A}). 
Therefore, we classify this object as a SB galaxy in this paper. 

\subsection{NGC 3628}
Starburst galaxy NGC 3628 ($L_{\rm 8-1000 \mu m}$ = 10$^{10.3}$ $L_\odot$; \citealt{2003AJ....126.1607S}) 
is a member of the Leo Triplet (Arp 317), including NGC 3627 and NGC 3623 as well. 
Radio observations of both continuum and hydrogen recombination line emissions 
revealed this galaxy shows a SB activity in its central $\sim$ 500 pc region (\citealt{1982ApJ...252..102C,1997ApJ...482..186Z}). 
A large scale galactic wind, relating this SB, 
has been detected in X-ray and H$\alpha$ emissions (\citealt{2001ApJ...560..707S,2004ApJ...606..829S}). 
In addition, a sub-kpc scale molecular outflow is also found (\citealt{2012ApJ...752...38T}). 
Note that an X-ray point source ($L_{\rm 0.3-8.0 keV}$ $\sim$ 10$^{40}$ erg s$^{-1}$), 
which is a candidate of an intermediate mass black hole, was discovered in this galaxy (\citealt{2001ApJ...560..707S}). 
But the location of the object is at least 20$''$ away from the nucleus. 
Thus, it would have no contribution to the molecular line observations with APEX 18$''$ beam 
used in this paper (\citealt{2014ApJ...784L..31Z}).  

\subsection{NGC 7552}
A nearly face-on barred spiral galaxy NGC 7552 ($L_{\rm 8-1000 \mu m}$ = 10$^{10.9}$ $L_\odot$; \citealt{2003AJ....126.1607S}) 
has been classified as a LINER galaxy due to its weak [OI] emission (\citealt{1988A&AS...75..273D}). 
It hosts a prominent circumnuclear ring with a radius of $\sim$ 2.5$''$ or 250 pc, 
which is the powering source of a SB activity (e.g., \citealt{1997ApJ...488..174S,2012A&A...543A..61B}). 
This SB ring is occupied by a number of young stellar clusters (\citealt{2012A&A...543A..61B}), 
which would be formed about 10 Myr ago (\citealt{1997ApJ...488..174S}). 
Inside the SB ring, neither X-ray (\citealt{2005ApJS..157...59L}) nor NIR (\citealt{1994ApJ...433L..13F}) 
observations have shown a clear evidence of a nuclear activity. 
Therefore, this galaxy is genuinely a SB galaxy. 
High resolution observations revealed that dense molecular gas is mainly 
concentrated to the SB ring (\citealt{2013ApJ...768...57P}). 

\subsection{IRAS13242-5713}
Unlike other sample galaxies, observational information on this LIRG 
($L_{\rm 8-1000 \mu m}$ = 10$^{11.0}$ $L_\odot$; \citealt{2003AJ....126.1607S}) is currently very limited. 
However, based on the MIR observations with {\it{Spitzer}} IRS, 
\citet{2012ApJ...744....2A} reported that an AGN, if it exists, has almost no contribution to the MIR spectrum. 
Indeed, [Ne V] 14.32 $\mu$m, which is an indicator of an existence of an AGN, was not detected (\citealt{2012ApJ...744....2A}). 
Then, although the observational evidence is lacked, we conclude that an AGN is absent, 
or at least negligible to the energy budget in this galaxy, 
and classify it as a SB galaxy. 

\subsection{N113}
This is an HII region in the Large Magellanic Cloud (LMC) associated with a clumpy molecular cloud (\citealt{2012ApJ...751...42S}) 
with a mass of a few $\times$ 10$^5$ $M_\odot$ in total (\citealt{2009ApJ...690..580W}). 
The molecular cloud is a site of massive star formation, 
as evidenced by several embedded young stellar objects (YSOs) (e.g., \citealt{2009ApJ...699..150S,2012ApJ...751...42S,2012A&A...542A..66C}). 
N113 hosts the most intense H$_2$O maser of the Magellanic Clouds 
(e.g., \citealt{2006MNRAS.372.1509O}) and OH maser as well (\citealt{1997MNRAS.291..395B}). 
This region (and the LMC as well) is characterized by its low-metallicity ISM ($Z$ $\sim$ 1/3 $Z_\odot$; e.g., \citealt{2009A&A...496..841H}). 
Therefore, this is an appropriate object to investigate the effect of metallicity on molecular chemistry. 

\subsection{N159}
This is also the best-studied HII region in the LMC, close to the evolved starburst of 30 Doradus. 
The N159 complex is a type-III giant molecular cloud as classified by \citet{2008ApJS..178...56F}, 
i.e., a GMC with HII regions and young star clusters. 
Indeed, this region is known to host many YSOs, 
OH maser sources, and ultra-compact HII regions (e.g., \citealt{2010ApJ...721.1206C}). 
Extensive studies of molecular material have been conducted for this region 
(e.g., \citealt{2000ApJ...545..234B,2010PASJ...62...51M,2011AJ....141...73M,2015A&A...580A..54O,2015arXiv151001246P}). 
The total mass of clumps visible at $^{12}$CO(3-2) emission is $\sim$ 6 $\times$ 10$^5$ $M_\odot$ (\citealt{2008ApJS..175..485M}). 
Recent non-LTE modelings of high density gas tracers such as HCN, HCO$^+$ and CS 
revealed the existence of warm ($T_{\rm kin}$ $\sim$ 80 K) 
and dense ($n_{\rm H_2}$ $\sim$ 5 $\times$ 10$^5$ cm$^{-3}$) gas (\citealt{2015arXiv151001246P}). 
As is the case of N113 (and LMC), this region is also a site with low-metallicity.

\section{B. Excitation of HCO$^+$(4-3) and CS(7-6) under non-LTE with photon trapping}\label{app-B}
\begin{figure*}
\epsscale{1}
\plotone{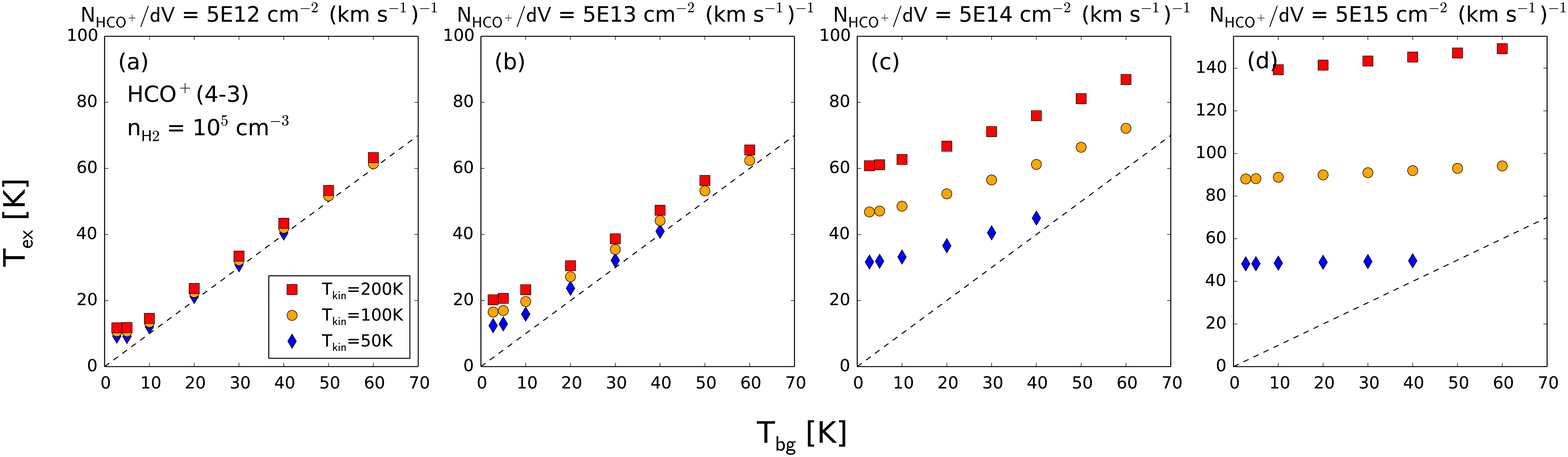}
\caption{Excitation temperature ($T_{\rm ex}$) of HCO$^+$(4-3) 
as a function of background temperature ($T_{\rm bg}$). 
The red square, orange circle, and blue diamond symbols indicate the models 
with gas kinetic temperature ($T_{\rm kin}$) of 200, 100, and 50 K, respectively. 
We here fixed gas volume density ($n_{\rm H_2}$) to 10$^5$ cm$^{-3}$. 
Four cases of the line-of-sight column density to the velocity width ratio ($N_{\rm HCO^+}/dV$) of 
(a) 5 $\times$ 10$^{12}$, (b) 5 $\times$ 10$^{13}$, (c) 5 $\times$ 10$^{14}$, and (d) 5 $\times$ 10$^{15}$ cm$^{-2}$ (km s$^{-1}$)$^{-1}$ are shown. 
Note that the scale of the $y$-axis in the panel-(d) is different from the others. 
The dashed line in each panel indicates the $T_{\rm ex}$ = $T_{\rm bg}$. 
Some cases where the line emission shows a maser feature are excluded. 
The overall trend is similar to that of HCN(4-3) as shown in Figure \ref{figure3}. 
}
\label{figure-App1}
\end{figure*}

\begin{figure*}
\epsscale{1}
\plotone{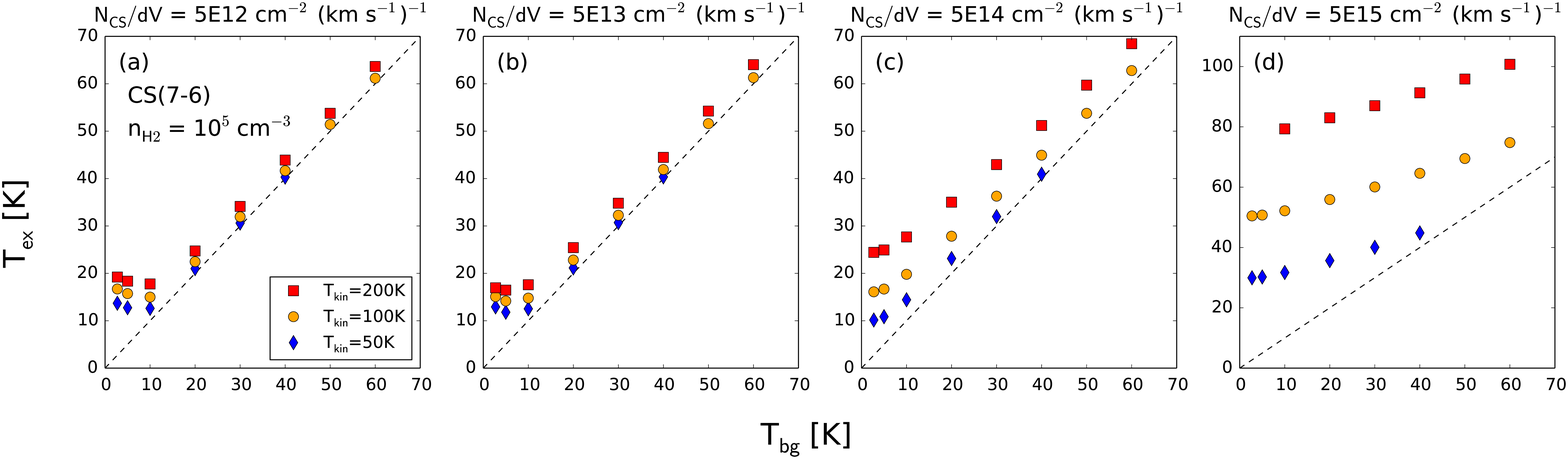}
\caption{Same as Figure \ref{figure-App1}, but for the cases of CS(7-6).}
\label{figure-App2}
\end{figure*}

The molecular excitation temperature ($T_{\rm ex}$) of HCO$^+$(4-3) and CS(7-6) are shown in Figures \ref{figure-App1} and \ref{figure-App2}, 
as a function of the background radiation temperature ($T_{\rm bg}$), respectively. 
The overall trend is totally similar to the case of HCN(4-3) as shown in Figure \ref{figure3}; 
$T_{\rm ex}$ approaches $T_{\rm bg}$ at $T_{\rm bg}$ $\ga$ 10 K, 
when the line is optically thin to moderately thick (left two panels). 
On the other hand, in the optically thick cases, $T_{\rm ex}$ is getting close to the gas kinetic temperature ($T_{\rm kin}$) 
and independent of $T_{\rm bg}$, due to an efficient photon trapping effect. 

\end{document}